\documentclass[aps,pra,twocolumn,showpacs,groupedaddress,superscriptaddress]{revtex4}

\usepackage{graphicx}

\begin{document}



\title{
	Nonlinear Magneto-Optical Rotation of Elliptically Polarized Light
}

\author{A.\ B.\ Matsko}
\affiliation{
	Department of Physics and Institute for Quantum Studies,
	Texas A\&M University, College Station, Texas 77843-4242
}

\author{I.\ Novikova}
\email{
	i.novikova@osa.org
}
\affiliation{
	Department of Physics and Institute for Quantum Studies,
	Texas A\&M University, College Station, Texas 77843-4242
}

\author{M.\ S.\ Zubairy}
\affiliation{
	Department of Physics and Institute for Quantum Studies,
	Texas A\&M University, College Station, Texas 77843-4242
}
\affiliation{
	Department of Electronics,
	Quaid-i-Azam University, Islamabad, Pakistan
}

\author{G.\ R.\ Welch}
\affiliation{
	Department of Physics and Institute for Quantum Studies,
	Texas A\&M University, College Station, Texas 77843-4242
}

\date{\today}

\begin{abstract}
We predict theoretically and demonstrate experimentally an
ellipticity-dependent nonlinear magneto-optic rotation of
elliptically-polarized light propagating in a medium with
atomic coherence.  We show that this effect results from
hexadecapole and higher order moments of the atomic coherence,
and is associated with an enhancement of Kerr and higher
orders nonlinearities accompanied by suppression of the other
linear and nonlinear susceptibility terms of the medium.  These
nonlinearities might be useful for quantum signal processing.
In particular, we report an observation of an enhancement of
the polarization rotation of elliptically polarized light
resonant with the $5S_{1/2} F=2 \rightarrow 5P_{1/2} F=1$
transition of ${}^{87}$Rb.
\end{abstract}


\maketitle

\section{Introduction}

	Nonlinear magneto-optic rotation (NMOR) of the
polarization plane of light resonant with atomic transitions is
attracting increasing attention~\cite{gawlik'94,budker_rmp'02}.
Ultranarrow (up to $1$~Hz~\cite{kanorsky'95,budker'98}) spectral
features accompanied by strong polarization rotation observed in
NMOR experiments are used (or proposed to be used) in sensitive
magnetometry~\cite{fleischhauer'00,budker'00a,novikova'02},
in time-reversal-invariance violation
experiments~\cite{hunter'91,bouchiat'95,budker_rev'99},
in measurements of the electron dipole
moment~\cite{barkov'88,kimball'01}, and in measurements of
various atomic constants~\cite{weis'01}.  Extremely slow
propagation of light has also been observed in NMOR in hot
rubidium vapor~\cite{budker'99a}.

	The most accurate description of the properties of NMOR
signals is obtained from an analysis of density matrix equations
for the atomic polarizations and populations along with Maxwell
equations describing propagation of the electromagnetic fields
in the atomic medium.  The exact solution of this problem,
however, is very complicated, and for most cases may be obtained
only numerically.  The problem should be somehow simplified
to obtain analytical results.

	The traditional approach to solution
of the problem is based on the approximation of
weak electromagnetic fields and low atomic vapor
densities~\cite{giraudcotton'82,drake'86,chen'90,kanorsky'93},
conditions found in early experiments involving incoherent
radiation from atomic discharge lamps.  In this case one
can use perturbation theory, and the atomic susceptibility
may be decomposed in a series of the electromagnetic fields
involved.  Magnetic field dependent terms of the susceptibility
decomposition which are nonlinear in the electromagnetic fields
are responsible for NMOR.  It can be demonstrated that only
two-photon processes are important in this approximation,
and therefore complicated multilevel systems may be reduced
to systems with small level number (such as $\Lambda$,
$V$, or $X$--schemes)~\cite{chen'90,junger'90,holmes'95}.
In this approximation, NMOR is a consequence of low frequency
ground-state coherence formed by two-photon processes between
Zeeman sublevels with difference in magnetic quantum numbers
equal to $\Delta m=\pm 2$.

	In some cases it is convenient to describe
the atom-light interaction from the point of view of
light-induced multipole moments of the atomic electron
distribution.  Conventionally this is done in terms
of an irreducible tensor representation of the density
matrix~\cite{d'yakonov'65,varshalovich_book,alexandrov_book}.
In this case, the ground-state coherence is equivalent to
the quadrupole moment, or alignment.  It has been suggested
that NMOR is a consequence of the alignment to orientation
conversion~\cite{budker'00}, where the orientation is equivalent
to the population difference between nearest Zeeman sublevels
with $\Delta m=\pm 2$.

	The simplified theoretical approaches used for
weak electromagnetic fields generally fail for strong ones.
The question that arises here is whether or not the interaction
with strong fields bring new physics, e.g.\ if the higher
order atomic coherences influence NMOR.  Alkali atoms have a
level structure which allows for a formation of the coherent
superposition of the magnetic sublevels with $\Delta m=\pm
4$ (hexadecapole moment in the multipole decomposition of
the interaction process) and even higher.  Such coherences
should be excited by multiphoton processes that include four
or more photons.  Gawlik \textit{et al.}~\cite{gawlik'74}
observed strong narrow features in a forward scattering
experiment with free sodium atoms, which were attributed
to a hexadecapole moment.  However, subsequent work of
Giraud-Cotton \textit{et al.}~\cite{giraudcotton'82} and other
groups~\cite{junger'90,chen'90,holmes'95} demonstrated that
these features may be explained using third-order perturbation
theory which includes only quadrupole moments.

	There have been a number of publications where
observation of hexadecapole and higher order moments is reported
for the case where the magnetic field is perpendicular to
the light propagation direction~\cite{theobald'89,suter'93}.
At the same time, the question of their influence on forward
scattering and NMOR signals in Faraday configuration is still
open~\cite{gawlik'96}.  Generally, the interpretation of the
experimental results in the case of strong laser fields and
large multipole moments is very complicated.  The high-order
coherence causes only slight modifications of the rotation
caused by the quadrupole moment, which hinders a convincing
demonstration of these high-order effects.

	We here solve both analytically and numerically the
problem of the propagation of strong elliptically polarized
electro-magnetic fields through resonant atomic media in the
presence of a magnetic field.  We particularly investigate
the properties of the light which interacts with the magnetic
sublevels in an $M$-like level configuration and, therefore,
forms coherences with $\Delta m =4$.  We demonstrate that
these coherences is responsible for a new type of polarization
rotation which depends on both the  light ellipticity and the
applied magnetic field. We observe this effect in hot vapor
of rubidium atoms.  Since such rotation does not appear for
an isolated $\Lambda$ scheme, our experiment may be treated
as a clear demonstration of the hexadecapole moment of atoms.

	Another interesting and important feature of the system
under consideration is connected with a large Kerr nonlinearity
that is associated with NMOR.  We analyze Kerr nonlinearity in
the $M$ level configuration and show that the ratio between
the nonlinearity and the absorption may be large.  Moreover,
we show that by increasing the number of Zeeman sublevels
(e.g.\ by using another Rb isotope or different alkali atom with
higher ground-state angular momentum) it is possible to realize
higher orders of nonlinearities.  Our method of creation of the
highly nonlinear medium with small absorption has prospects
in fundamental as well as applied physics.  It can be used
for construction of nonclassical states of light as well as
coherent processing of quantum information~\cite{qq1}.

	To bridge between this and previous studies we should
note that NMOR may be attributed to coherent population
trapping (CPT)~\cite{CPT,sz} and electromagnetically
induced transparency (EIT)~\cite{EIT}.  Both EIT and
CPT are able to suppress linear absorption of resonant
multilevel media while preserving a high level of nonlinear
susceptibility~\cite{harris90prl,largenl,luknature}.
Previous theoretical studies of coherent media with large
optical Kerr nonlinearities have described nonlinearities
resulting from the effective self-action of an electromagnetic
field at a single photon energy level, such as a photon
blockade~\cite{imamoglu'97,rebic'99,gheri'99,greentree'00}, or
an effective interaction between two electromagnetic fields due
to refractive~\cite{largenl,luknature,harris99prl,lukin00prl}
and absorptive~\cite{absorption} Kerr nonlinearities.
The absorptive $\chi^{(3)}$ nonlinearities
were studied experimentally for quasiclassical
cases~\cite{hemmer_abs,absorption_exp}.  It was shown quite
recently, that a similar approach may lead to achievement of
even higher orders of nonlinearity~\cite{zubairy02pra}.

\begin{figure}
\centering
\includegraphics[width=1.00\columnwidth]{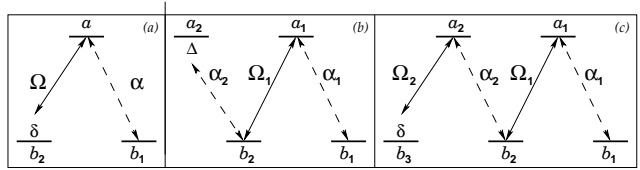}
\caption{\label{fig1.fig}
    Energy level schemes for
    (a) $\Lambda$-system;
    (b) $N$-system;
    (c) $M$-system.
}
\end{figure}

	A method of producing Kerr nonlinearity with vanishing
absorption is based on the coherent properties of a three-level
$\Lambda$ configuration (see Fig.~\ref{fig1.fig}a).  In such
a scheme the effect of EIT can be observed.  Two optical
fields, $\alpha_1$ and $\Omega_1$, resonant with the
transitions of the $\Lambda$ system, propagates through
the medium without absorption.  However, because an ideal
EIT medium does not interact with the light, it also can
not lead to any nonlinear effects at the point of exact
transparency~\cite{sz}.  To get a nonlinear interaction
in the coherent medium one needs to ``disturb'' the EIT
regime by introducing, for example, additional off-resonant
level(s) (level $a_2$ in Fig.~\ref{fig1.fig}b).  In the
following we refer to the resultant level configuration
an $N$-type scheme.  Such a scheme has been used in previous
works~\cite{largenl,luknature,imamoglu'97,rebic'99,gheri'99,greentree'00,lukin00prl}.
If the disturbance of EIT is small, i.e., the detuning $\Delta$
is large, the absorption does not increase significantly.
At the same time, the nonlinearity can be as strong as the
nonlinearity in a near-resonant two level system.

	This paper is based on the existence of CPT
in multilevel media.  Unlike the early ideas of Kerr
nonlinearity enhancement, we propose to use not a single
$\Lambda$ scheme, but several coupled $\Lambda$ schemes.
In particular, we consider the $M$-type configuration as shown
in Fig.~\ref{fig1.fig}c.  Coherent population trapping exists
in such a scheme, like in a $\Lambda$-type level system.

	By introducing a small detuning, $\delta$,
we may disturb this CPT and produce a strong nonlinear
coupling among the electromagnetic fields interacting
with the atomic system, while having small absorption of
the fields~\cite{matsko_prep'02}.  The dispersion of the $M$
level media and associated group velocity of light propagating
in the media are intensity dependent due to the nonlinearity,
as was theoretically predicted by A.~Greentree~\textit{et
al.}~\cite{greentree'02prep}.  Finally, in the case discussed
below, energy levels of the $M$ configuration correspond to
Zeeman sublevels of alkali atoms.  The multi-photon detuning
is introduced by a magnetic field, resulting in the intensity
dependent polarization rotation.

	We show a simple way to reduce a five-level $M$
configuration to a four-level $N$ configuration, and
prove that these completely different schemes demonstrate
refractive nonlinearities of the same magnitude.  This is a
very interesting result, because the nonlinearity of the $M$
configuration is a consequence of the hexadecapole part
of atomic coherence, while the nonlinearity in the $N$
configuration results from quadrupole atomic coherence.

	Our paper is organized as follows.  In Sec.~II
we analyze the $F=1 \rightarrow F'=0$ atomic transition,
demonstrate that this transition may be described by a $\Lambda$
level configuration, and show that the polarization rotation
in the case of a $\Lambda$ configuration does not depend on
the light ellipticity.  In Sec.~III we study $F=2 \rightarrow
F'=1$ atomic transitions, show that it consists of $\Lambda$
and $M$ schemes, investigate properties of the $M$ interaction
scheme, and show that ellipticity dependent NMOR is possible.
Using analytical calculations we show that the hexadecapole
moment plays an important role here.  In Sec.~IV we expand
our theory to the case of generalized $M$ energy level systems
and discuss possibilities of observations of $\chi^{(5)}$ and
higher order nonlinearities.  In Sec.~V we discuss applications
of the nonlinearities for quantum information processing.
The case of Doppler broadened $\Lambda$, $M$ and $N$ systems
is considered in Sec.~VI for the particular case of a weak
probe field.  We present experimental measurements of the
polarization dependent NMOR in hot Rb$^{87}$ and Rb$^{85}$
atomic vapors in Sec.~VII.  Finally, in Sec.~VIII, we present
our conclusions based on these results.

\section{Analysis of NMOR for the case of an $F=1 \rightarrow
F'=0$ transition}

\begin{figure}
\centering
\includegraphics[width=1.00\columnwidth]{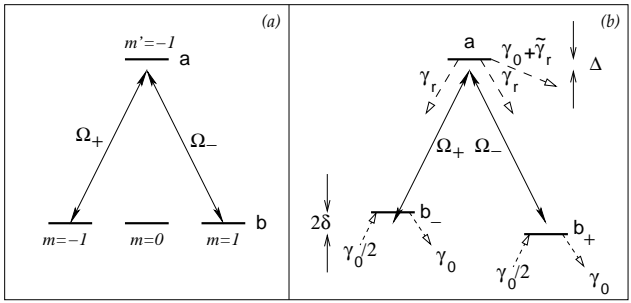}
\caption{\label{lambda}
	(a) Schematic interaction of an electromagnetic wave
	with an atomic transition $|b\rangle,\ F=1 \rightarrow
	|a\rangle,\ F'=0$.  The electromagnetic field is
	decomposed into two circularly polarized components
	having Rabi frequencies $\Omega_+$ and $\Omega_-$.
	(b) Simplification of the scheme (a) for the case when
	there is a magnetic field applied parallel to the wave
	vector of the electromagnetic wave.
}
\end{figure}

	A three-level $\Lambda$ configuration is the simplest
system that results in NMOR.  This system appears naturally in
the configuration of Zeeman sublevels of an $F=1 \rightarrow
F'=0$ atomic transition, where $F$ and $F'$ are the total
angular momenta of the ground and excited atomic states,
respectively.  This scheme can be easily seen if the angular
momentum quantization axis is chosen along the light propagation
direction.  The effective interaction scheme for this case
is shown in Fig.~\ref{lambda}a.  The $\Lambda$ configuration
consists of two circularly polarized components of the laser
field,  which create the low-frequency coherence between
magnetic sublevels $m=\pm 1$. Because of the selection rules,
the electromagnetic waves do not interact with the sublevel
having $m=0$.

	For zero magnetic field such a configuration
demonstrates coherent population trapping.  A nonzero magnetic
field collinear with the wave vector of the light leads to Zeeman
shift of magnetic sublevels $m=\pm 1$, which disturbs CPT and
results in an interaction between the light and the atoms.
The nonlinear polarization rotation emerges as a consequence
of this interaction.

	In the following, we briefly review the basic properties
of CPT in $\Lambda$ systems and calculate the optical losses
and the polarization rotation by solving the optical Bloch
equations for the density matrix elements.  Finally, we note
how the $F=1 \rightarrow F'=0$ level configuration can be
reduced to a $\Lambda$ system via proper renormalization of
decay rates and density matrix.

\subsection{Coherent population trapping in a $\Lambda$ system}

	The Hamiltonian for the $\Lambda$ system shown in
Fig.~\ref{lambda}b can be written as
\begin{eqnarray}
H_\Lambda &=& \hbar \Delta |a\rangle \langle a| -\hbar \delta
|b_+\rangle \langle b_+| + \hbar
\delta |b_-\rangle \langle b_-| \nonumber \\
&+& \hbar \left( \Omega_-|a\rangle \langle b_+| +
\Omega_+|a\rangle \langle b_-| + H.c.\ \right) \label{ham-Lambda1}
\end{eqnarray}
where $E_+$ and $E_-$ are the electric field amplitudes of two
opposite circularly polarized electromagnetic waves, $\Omega_-
= E_- \wp_{ab+}/\hbar$, $\Omega_+ = E_+ \wp_{ab-}/\hbar$
are the corresponding complex Rabi frequencies, $\wp_{ab+}$
and $\wp_{ab-}$ are the atomic dipole moments, $\Delta$
is the one-photon detuning of the laser frequency from the
exact atomic transition, and $\delta$ is the shift of the
ground-state sublevels resulting, for example, from interaction
with a magnetic field.

	The eigenvalues of this Hamiltonian $\lambda_i$
(where $H |\lambda \rangle = \hbar \lambda |\lambda \rangle$) may
be found from
\begin{eqnarray} \label{ei1}
 \left | \matrix  {
   \delta-\lambda & \Omega_+^* & 0 \cr
   \Omega_+ & \Delta-\lambda & \Omega_- \cr
   0 & \Omega_-^* & -\lambda - \delta } \right | = 0
\end{eqnarray}
or
\begin{eqnarray} \label{ei1-eq}
-\lambda^3+\lambda^2 \Delta+\lambda(\delta^2+|\Omega_+|^2+|\Omega_-|^2) - \\
\nonumber \delta (\delta \Delta + |\Omega_-|^2-|\Omega_+|^2)=0~.
\end{eqnarray}

	In the degenerate case ($\delta =0$) the eigenvalues
and corresponding eigenstates are
%
%
\begin{eqnarray} \label{Lambda-dark}
\lambda_D &=& 0 \nonumber \\
|D\rangle &=& \frac{\Omega_+ |b_+\rangle -
\Omega_- |b_-\rangle}{\sqrt{|\Omega_+|^2+|\Omega_-|^2}} \\
\lambda_{B_{1,2}} &=& \frac{\Delta}{2} \pm
\sqrt{\frac{\Delta^2}{4}+ |\Omega_+|^2+|\Omega_-|^2}
\nonumber \\
|B_{1,2}\rangle &=& \sqrt{\frac{| \lambda_{B_{1,2}}|}{
\lambda_{B_{1}}- \lambda_{B_{2}}}}  \left ( |a\rangle +
\frac{\Omega_+^*}{\lambda_{B_{1,2}}}
 |b_-\rangle + \frac{\Omega_-^*}{\lambda_{B_{1,2}}}
 |b_+\rangle \right )\,. \nonumber \\
 &~&
\end{eqnarray}
The state denoted as $|D\rangle$ is called the ``dark state''
because an atom in this state does not interact with the light
fields and, therefore, does not fluoresce.  Atoms in the other
two states, called ``bright states'', readily absorb light.
Therefore, atoms initially prepared in a bright state are
optically pumped into the dark state after some finite time
comparable with the lifetime of the excited level $|a\rangle$.
Thus, in steady-state, the atomic ensemble does not interact
with the electro-magnetic fields, which is the essence of CPT.
The dispersive properties of the atomic system in the dark
state are governed by the coherence between the ground states
of the $\Lambda$ system.  The corresponding density matrix
element may be found from (see Ref.~\ref{Lambda-dark}):
\begin{equation} \label{coh1}
\rho_{b+b-} = -\frac{\Omega_-^* \Omega_+}{|\Omega_-|^2+ |\Omega_+|^2}~.
\end{equation}

	The true dark state exists only for $\delta = 0$.
As soon as the exact resonant conditions are disturbed, the
system starts interacting with light.  However, for small
detunings ($\sqrt {|\Omega_+|^2+|\Omega_-|^2} \gg |\delta|,\
\sqrt{|\Delta \delta|}$) the disturbance of the dark state
is small, and most of the atomic population is concentrated
in the modified dark state $|\tilde{D}\rangle$.  In this case
the eigenvalue $\tilde \lambda_D$ corresponding to this state
can be found by solving Eq.~(\ref{ei1-eq}) and keeping only
the terms linear in $\delta$:
\begin{eqnarray} \label{eigen-Lam}
\tilde{\lambda}_D &=& \delta
\frac{|\Omega_-|^2-|\Omega_+|^2} {|\Omega_+|^2+|\Omega_-|^2} \\
\label{eigen-vec-dark}
|\tilde{D}\rangle &\simeq&  \mathcal {N}\left\{ |D\rangle + 2\delta
\frac{\Omega_+\Omega_-}{(|\Omega_+|^2+|\Omega_-|^2)^{3/2}}|a\rangle \right \}
\end{eqnarray}
where $\mathcal{N}\simeq 1+O(\delta^2)$ is a normalization
constant.  From Eq.~(\ref{eigen-vec-dark}) it is obvious that
the population of the excited level $|a\rangle$ is proportional
to $\delta^2$.

\subsection{Equations of motion}

	It is possible to obtain the equation of motion for
the electro-magnetic fields, using the method reported in
Ref.~\cite{zubairy02pra,matsko_prep'02}.  If we assume a small
disturbance of CPT, almost all atomic population remains in a
dark state during the interaction process, and we can rewrite
the interaction Hamiltonian as
\begin{equation}
H \simeq \hbar \tilde{\lambda}_D|\tilde{D}\rangle
\langle\tilde{D}|~.
\end{equation}
Since now $|\tilde{D}\rangle \langle\tilde{D}| \simeq 1$, the
atomic degrees of freedom may be excluded from the interaction
picture and we can write $H \simeq \hbar \tilde{\lambda}_D$\,.
The interaction Hamiltonian may be rewritten in the
Heisenberg picture, so that $\Omega \propto \hat a$, where
$\hat a$ is the annihilation operator for the electromagnetic
field~\cite{zubairy02pra}.  The quantum mechanical equation
for the electromagnetic creation and annihilation operators
may be presented in the following form:
\begin{equation} \label{quant}
\frac{d\hat a}{dt} = -\frac{i}{\hbar} \frac{\partial H}{\partial
\hat a^\dag}~.
\end{equation}
The propagation equation for the electromagnetic field
amplitude $E$ can be obtained from Eq.~(\ref{eigen-Lam}) as a
quasiclassical analogue of Eq.~(\ref{quant})~\cite{boyd_book}:
\begin{equation} \label{e-prop}
\frac{\partial E}{\partial z} = 2\pi i N \frac{\nu}{c}\frac{\partial H}
{\partial E^*}
\end{equation}
where $N$ is the density of the atoms in the cell, and
$\nu$ is a carrier frequency of the electromagnetic wave.
Using Eqs.~(\ref{e-prop}) and (\ref{eigen-Lam}) (with $H
\simeq \hbar \tilde{\lambda}_D$) we arrive at the following
propagation equations for the Rabi frequencies $\Omega_\pm$:
\begin{equation} \label{e-of-m_Lambda}
\frac{\partial \Omega_{\pm }}{\partial z} = \mp 2i \kappa
\delta \Omega_{\pm}
\frac{|\Omega_{\mp}|^2} {(|\Omega_+|^2+|\Omega_-|^2)^2}
\end{equation}
where $\kappa$ is a coupling constant given by
\begin{equation} \label{kappa}
\kappa  = \frac{3}{8\pi} N \lambda^2 \gamma_r
\end{equation}
and $\lambda$ is the wavelength of the light in vacuum.
It is also useful to rewrite the equation of motion for the
field amplitudes $E_\pm$:
\begin{eqnarray}
 \frac{\partial E_\pm}{\partial z}& = &\mp 4 i \pi \hbar \delta N \frac
{\nu}{c} E_\pm \frac{|E_\mp|^2}{(|E_{+}|^2+|E_{-}|^2)^2}~.
\label{e1a}
\end{eqnarray}

	Equation~(\ref{e-of-m_Lambda}) is suitable for
describing the phase evolution of the electromagnetic fields.
However the decay processes responsible for the optical
losses cannot be correctly included in this method and we
need a density matrix approach.  In the following section
we explicitly calculate the density matrix elements for the
$\Lambda$ system to verify Eq.~(\ref{e-of-m_Lambda}) and
discuss the attenuation of the light.

\subsection{Density matrix approach}

	In order to discuss a realistic model of the atom-field
interaction in an atomic cell we need to include atomic level
decay rates (Fig.~\ref{lambda}b).  We introduce the decay rate
$\gamma_0$ outside of the system that is inversely proportional
to the finite interaction time of the atoms and electromagnetic
field.  This decay represent the atoms leaving the interaction
region.  Another term that describes decay to outside levels,
$\tilde\gamma_r$, describes population pumping into states
that do not interact with the fields, for example, the state
with zero magnetic moment ($m=0$ in Fig.~\ref{lambda}a).
The natural decay rate from level $|a\rangle$ to levels
$|b_+\rangle$ or $|b_-\rangle$ is denoted as $\gamma_r$.

	We also need to take into account the atoms entering the
laser beam.  To do that, we include incoherent pumping to all
Zeeman sublevels from outside of the system, which means that
atoms that enter the interaction region have equal populations
of the ground state sublevels and no coherence between them.
The value of the incoherent pumping rate is chosen to be
$\gamma_0/2$ to keep the sum of level populations equal to unity
in the case of $\tilde\gamma_r =0$.  When $\tilde\gamma_r \ne
0$, the sum of the populations is less then unity because of
the optical pumping, i.e.,
\begin{equation} \label{norml}
\rho_{aa} + \rho_{b+b+} + \rho_{b-b-}=1- \frac{\tilde
\gamma_r}{\gamma_0} \rho_{aa}~.
\end{equation}

	The time-evolution equations for the density matrix
elements $\rho_{ij}$ for the $\Lambda$ system can be obtained
from the Liouville equation:
\begin{equation}
\dot{\rho} = -\frac{i}{\hbar}\left[ H_\Lambda,\rho\right] -
\frac12 \{\Gamma,\rho\} + R
\end{equation}
where $\rho = \sum \rho_{ij} |i\rangle \langle j|$, $H_\Lambda$
is given by Eq.~(\ref{ham-Lambda1}), $\Gamma$ is the matrix
describing the decays in the system, and $R$ is the matrix of
incoherent pumping to the ground state sublevels.  Then the
equations for the atomic populations are:
\begin{eqnarray}
\dot \rho_{b-\,  b-} &=&  \frac{\gamma_0}{2} - \gamma_0 \rho_{b-\,
b-}
+ \gamma_{ r} \rho_{a\, a}\nonumber\\
&& +i(\Omega_+^*\rho_{a\, b-}-c.c.) \label{sb-b-}\\
\dot \rho_{b+\,  b+} &=&  \frac{\gamma_0}{2} - \gamma_0 \rho_{b+\,
b+}
+ \gamma_{ r} \rho_{a\, a}\nonumber\\
&& +i(\Omega_-^*\rho_{a\, b+}-c.c.)\,.  \label{sb+b+}
\end{eqnarray}
Analogously, for the polarizations we have
\begin{eqnarray}
\dot \rho_{a \, b\pm} &=& -\Gamma_{a \, b\pm} \rho_{a \, b\pm} +
i\Omega_{\mp} (\rho_{b\pm \, b\pm}-\rho_{a\,a}) \nonumber\\
&& +i \Omega_{\pm} \rho_{b \mp \, b \pm} \label{sab+-}\\
\dot \rho_{b- \, b+} &=& -\Gamma_{b- \, b+} \rho_{b- \, b+} +
i\Omega_+^* \rho_{a \, b_+} \nonumber\\
&& - i \Omega_- \rho_{b- \, a} \label{sbb+-}
\end{eqnarray}
where
\begin{eqnarray}
\Gamma_{a \, b\pm} &=& \gamma +
i \left (\Delta \pm \delta  \right ) \\
 \Gamma_{b- \, b+} &=& \gamma_0 + 2i\delta
\end{eqnarray}
with $\gamma = \gamma_r + \gamma_0 + \tilde \gamma_r/2$.

	In the steady state case, we can solve
Eqs.~(\ref{sab+-}) and (\ref{sbb+-}) in terms of the atomic
populations:
%
%
\begin{eqnarray} \label{sbb-+1}
\rho_{b- \, b+} &=& - \frac{ \Omega_+^*\Omega_- \left (
\displaystyle \frac{n_{b- \, a}}{\Gamma_{b_- \, a}} +
\displaystyle \frac{n_{b+ \, a}}{\Gamma_{a \, b_+}} \right ) }
{\Gamma_{b- \, b+} + \displaystyle \frac
{|\Omega_+|^2}{\Gamma_{a\, b+}} + \displaystyle \frac
{|\Omega_-|^2}{\Gamma_{b- \, a}}
} \\
\rho_{a \, b\pm} &=& \frac{i\Omega_\mp}{\Gamma_{a\, b\pm}}
\displaystyle \frac{ n_{b\pm\, a} \left ( \Gamma_{b\mp\, b\pm} +
\displaystyle \frac{|\Omega_\mp|^2}{\Gamma_{b\mp\, a}} \right )-
n_{b\mp\, a} \displaystyle \frac{|\Omega_\pm|^2} {\Gamma_{b\mp\,
a}} } {\Gamma_{b\mp \, b\pm} + \displaystyle \frac
{|\Omega_\pm|^2}{\Gamma_{a\, b\pm}} + \displaystyle \frac
{|\Omega_\mp|^2}{\Gamma_{b\mp \, a}}} \nonumber\\
&&
\label{rhoabpm}
\end{eqnarray}
where $n_{b\pm\, a} \equiv \rho_{b\pm\, b\pm} -\rho_{aa}$\,.
Inserting these expressions into Eqs.~(\ref{sb-b-}) and
(\ref{sb+b+}) and using the condition given in Eq.~(\ref{norml})
we can derive linear equations for the atomic populations.
In the general case, however, their solution is very cumbersome.

	Let us consider the case of a strong electro-magnetic
field, such that $|\Omega|^2/\gamma_0 \gamma \gg 1$.  We also
assume that $|\delta|, \gamma_0 \ll \gamma, |\Omega|$,
and $\Delta = 0$.  In the zeroth approximation the atomic
populations are determined by Eq.~(\ref{eigen-Lam}):
\begin{eqnarray}
\rho_{b\pm\, b\pm}^{(0)} &\simeq& \frac{|\Omega_\pm|^2}{|\Omega|^2} \\
\rho_{a\, a}^{(0)} &\simeq& 0
\end{eqnarray}
where $|\Omega|^2 = |\Omega_+|^2 +|\Omega_-|^2$.

	Now we can solve for the polarizations $\rho_{a\,
b\pm}$\,, keeping only the terms linear in $\delta$ and
$\gamma_0$
\begin{equation} \label{polar-fin}
\rho_{a\, b\pm} \simeq \frac{i\Omega_{\mp}} {|\Omega|^4} \left
(\frac{\gamma_0}{2}|\Omega|^2 \pm 2 i \delta|\Omega_{\pm}|^2
\right )~.
\end{equation}
It is important to note that this expression for the
polarization, obtained for an open $\Lambda$ system, coincides
with the analogous expression calculated by Fleischhauer
\textit{et al.}~\cite{fleischhauer'00} for a closed system,
if the ground-state coherence decay rate and the population
exchange rate between ground states are the same and equal
to $\gamma_0$.  This proves the equivalence of the open and
closed models for the description of $\Lambda$ schemes,
which has been previously demonstrated by Lee \textit{et
al.}~\cite{lee_prep'02} for the particular case of a weak
probe field.

	The stationary propagation of two circularly polarized
components of the laser field through the atomic medium is
described by Maxwell-Bloch equations for the slowly-varying
amplitudes and phases:
\begin{equation} \label{Lambda-e1}
\frac{\partial\Omega_{\pm}}{\partial z} \simeq -
\kappa\frac{\Omega_{\pm}} {|\Omega|^4} \left (
\frac{\gamma_0}{2}|\Omega|^2 \pm 2i \delta|\Omega_{\mp}|^2 \right
).
\end{equation}
Note that Eq.~(\ref{e-of-m_Lambda}) can be obtained from
Eq.~(\ref{Lambda-e1}) in the limit $\gamma_0 = 0$.

	Separating the real and imaginary parts
of Eq.~(\ref{Lambda-e1}) and using $\Omega_{\pm} =
|\Omega_{\pm}|~e^{i\phi_\pm}$\,, one can find the propagation
equations of the electromagnetic field intensity $|\Omega|^2$
and the rotation angle of the polarization ellipse
$\phi=(\phi_+-\phi_-)/2$:
\begin{eqnarray} \label{lambda_prop}
\frac{\partial|\Omega|^2}{\partial z} &=& -\kappa \gamma_0 \\
\frac{\partial\phi}{\partial z} &=&
-\frac{2\kappa\delta}{|\Omega|^2}.
\end{eqnarray}
After integration, the following expressions for the light
transmission $I_{\mathrm{out}}$ and the polarization rotation angle
$\phi$ are obtained:
\begin{eqnarray}
I_{\mathrm{out}} &=& I_{in} \left( 1- \frac{\kappa
\gamma_0 L}{|\Omega(0)|^2}
\right) \label{lambda-transm} \\
\phi &=& \frac{2\delta}{\gamma_0} \ln \frac {I_{in}}{I_{out}}
\label{lambda-rot}
\end{eqnarray}
where $L$ is the interaction length.  It is important to note
that the final expressions in Eqs.~(\ref{lambda-transm}) and
(\ref{lambda-rot}) include only the total laser intensity,
not the intensities of the individual circular components.
This means that both transmission and polarization rotation are
independent of the initial polarization of light~\cite{note1}.

\subsection{Normalization conditions for an $F=1 \rightarrow F'=0$
transition}

	The correspondence between the $F=1 \rightarrow
F'=0$ scheme (Figs.~\ref{lambda}a) and $\Lambda$ scheme
(Fig.~\ref{lambda}b) can be obtained if we exchange $\gamma_r$
by $\gamma_{aa}/3$, where $\gamma_{aa}$ is the decay rate
of the excited state to the ground state.  The decay rate
$\tilde \gamma_r$ should be presented as $\tilde \gamma_r =
\gamma_{aa}/3 + \tilde \gamma_{aa}$, where $\tilde \gamma_{aa}$
stands for the decay of the excited state outside of the system
in Fig.~\ref{lambda}a.

	We assume that the incoherent pumping rate into each
Zeeman ground state is equal to $\gamma_0/3$, to keep the
normalization condition similar to Eq.~(\ref{norml}):
\begin{equation} \label{norm10}
\tilde \rho_{aa} + \rho_{+1,+1} + \rho_{-1,-1} +\rho_{0,0} = 1-
\frac{\tilde \gamma_{aa}}{\gamma_0} \tilde \rho_{aa}
\end{equation}
where $\tilde \rho_{aa}$ is the population of the excited
state and $\rho_{ii}$ is the population of the $i$th magnetic
sublevel of the ground state in the system depicted in
Fig.~\ref{lambda}a.

	Keeping in mind that the population of the $m=0$ state
is determined by the decay rate of excited state $|a\rangle$ and
by the decay outside of the system we write the rate equation
\begin{equation} \label{rho00}
\dot \rho_{0,0} = \frac{\gamma_0}{3} - \gamma_0 \rho_{0,0} +
\frac{\gamma_{aa}}{3} \tilde \rho_{aa}
\end{equation}
and solve it in the steady state
\begin{equation} \label{xi2}
\rho_{0,0} = \frac{1}{3}+\frac{\gamma_{aa}}{3\gamma_0}\tilde
\rho_{aa}~.
\end{equation}

	Let us assume that $\tilde \rho_{aa} = \xi \rho_{aa}$,
$\rho_{+1,+1} = \xi \rho_{b+b+}$, and $\rho_{-1,-1} = \xi
\rho_{b-b-}$.  The normalization parameter $\xi$ can be found
by substituting Eq.~(\ref{xi2}) into Eq.~(\ref{norm10}), and
comparing the normalization conditions Eqs.~(\ref{norml}) and
(\ref{norm10}):
\begin{equation} \label{xi3}
\xi = \frac{2}{3}.
\end{equation}
Therefore, we can derive density matrix elements for the $F=1
\rightarrow F'=0$ level scheme shown in Fig.~\ref{lambda}a by
simple multiplication of the elements of the density matrix
for the $\Lambda$ scheme by the scaling factor $\xi$.

\section{Analysis of NMOR for the case of an $F=2\rightarrow
F'=1$ transition}

	For atomic ground atomic state angular momentum higher
than $F=1$ it is possible to create more than one $\Lambda$
link between magnetic sublevels.  This is equivalent to the
creation of coherent atomic states characterized by higher
angular momenta, which may drastically change the interaction
of such a medium with the electromagnetic field.

\begin{figure}
\centering
\includegraphics[width=1.00\columnwidth]{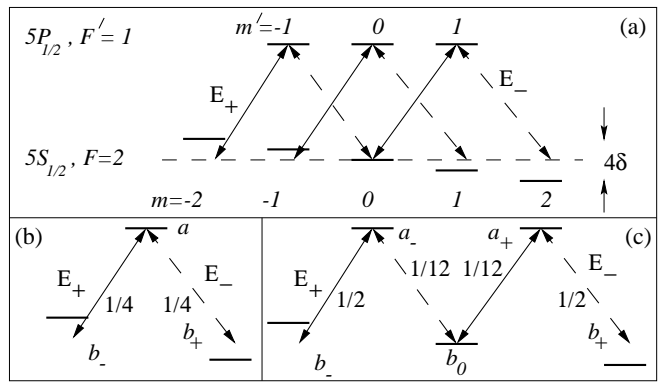}
\caption{\label{levels21.fig}
	a) Energy level scheme for $^{87}$Rb atoms.
	This scheme may be decomposed into a superposition
	of b) $\Lambda$-system and c) $M$-system. Transition
	probabilities are shown for each individual transition.
}
\end{figure}

	Let us concentrate first on $F=2 \rightarrow F'=1$
transitions, which occur in the ${}^{87}$Rb $\mathrm{D}_1$
line.  The case of higher angular momenta is discussed in the
next section.  Interaction of elliptically polarized light
with the $F=2 \rightarrow F'=1$ transition may be decomposed
into a $\Lambda$ scheme with $m=-1 \leftrightarrow m'=0
\leftrightarrow m=+1$, and an $M$ scheme $m=-2 \leftrightarrow
m'=-1 \leftrightarrow m=0 \leftrightarrow m'=+1 \leftrightarrow
m=+2$, as shown in Fig.~\ref{levels21.fig}a.  The main
difference of between an $M$ scheme and a $\Lambda$ scheme
is that the higher order coherence ($\Delta m=4$) becomes
important.  Since the $\Lambda$ system had been studied in the
previous section, we primarily concentrate on $M$ scheme here.

	The $M$ scheme is described by a set of twelve
density matrix equations.  The only straightforward way to
solve this system is with numerical methods.  However, if
we study the atomic interactions with weak magnetic fields,
the decay processes and polarization rotation processes are
independent, as we saw for the $\Lambda$ configuration.  Thus,
the polarization rotation may be found in analytical form
under the condition of zero relaxation using the Hamiltonian
diagonalization procedure as presented for the $\Lambda$
system.  The modified Schr\"odinger equation model is suited
for this as well.  The optical losses may be found separately
by considering the optical pumping into the dark state with
zero magnetic field.

\subsection{Coherent population trapping in an $M$ level scheme}

	It has been shown that the dark state
exists even for atoms with complicated Zeeman
substructure interacting with elliptically polarized
light~\cite{yudin'89,yudin'96,yudin'98,prior'98,varcoe'99,wang'00}.
Here we recall the analytical expressions for this dark
state and the corresponding eigenvalues.  Using an effective
interaction Hamiltonian, we derive propagation equations for
the electromagnetic fields.  We restrict our consideration
to the case relevant to the $M$ configuration consisting of
Zeeman energy sublevels in the magnetic field.  That is,
we assume that the atomic transition frequencies are such
that $\omega_{a-b0}= \omega_{a+b0} = \omega$, $\omega_{a-b-}
= \omega-2\delta$, and $\omega_{a+b+} = \omega+2\delta$,
where the detuning $\delta$ is due to a Zeeman shift, and the
laser frequency $\nu$ is resonant with the atomic transition.
The interaction Hamiltonians for $M$ systems is
\begin{eqnarray}
H_M &=& -2\hbar \delta |b_+\rangle \langle b_+| + 2\hbar \delta
|b_-\rangle \langle b_-| \nonumber \\*
 &+& \hbar (
\Omega_{1-}|a_+\rangle \langle b_+| + \Omega_{2-}|a_-\rangle
\langle b_0| \nonumber \\*
 &+& \Omega_{1+}|a_+\rangle \langle b_0| +
\Omega_{2+}|a_-\rangle \langle b_-| + H.c. ) \label{ham-M1},
\end{eqnarray}
where $\Omega_{1-} = E_- \wp_{a+b+}/\hbar$, $\Omega_{1+} = E_+
\wp_{a+b0}/\hbar$, $\Omega_{2-} = E_- \wp_{a-b0}/\hbar$,
$\Omega_{2+} = E_+ \wp_{a-b-}/\hbar$ (see
Fig.~\ref{levels21.fig}c).

	As in the $\Lambda$ system, the eigenvalues of the
interaction Hamiltonian can be determined from:
\begin{eqnarray} \label{ei2}
&& \left | \matrix  {
   2\delta-\lambda & \Omega_{2+}^* & 0 & 0 & 0 \cr
   \Omega_{2+} & -\lambda & \Omega_{2-} & 0 & 0 \cr
   0 & \Omega_{2-}^* & -\lambda & \Omega_{1+}^* & 0 \cr
   0 & 0 & \Omega_{1+} & -\lambda & \Omega_{1-} \cr
   0 & 0 & 0 & \Omega_{1-}^* & -\lambda - 2 \delta} \right | = 0
\end{eqnarray}
Also, as in the $\Lambda$ scheme, the eigenvalue $\lambda=0$
and corresponding ``dark state'' exists only for $\delta = 0$:
\begin{equation} \label{M-dark}
|D\rangle = \frac{\Omega_{1+}\Omega_{2+} |b_+\rangle -
\Omega_{1-}\Omega_{2+} |b_0\rangle +
\Omega_{1-}\Omega_{2-}|b_-\rangle}{\sqrt{|\Omega_{1-}|^2|\Omega_{2+}|^2
+ |\Omega_{1+}|^2|\Omega_{2+}|^2 +
|\Omega_{1-}|^2|\Omega_{2-}|^2}}
\end{equation}
It is worth noting that, similar to the $\Lambda$ system,
the non-vanishing low frequency coherences $\rho_{b+b0}$ and
$\rho_{b0b-}$ are important here.  The major difference in the
dispersive properties of the $M$ and $\Lambda$ schemes arises
from the existence of the four-photon coherence $\rho_{b+b-}$:
\begin{equation} \label{coh2}
\rho_{b+b-} = \frac{\Omega_{1-}^*\Omega_{2-}^* \Omega_{1+}\Omega_{2+}}
{|\Omega_{1-}|^2|\Omega_{2+}|^2
+ |\Omega_{1+}|^2|\Omega_{2+}|^2 +
|\Omega_{1-}|^2|\Omega_{2-}|^2}~.
\end{equation}
For small $\delta$ we can again find the eigenvalue for the
quasi-dark state, taking into account only the linear terms
in $\delta$:
\begin{equation} \label{eigen-M}
\tilde{\lambda}_{M} = 2\delta
\frac{|\Omega_{2-}|^2|\Omega_{1-}|^2-
|\Omega_{2+}|^2|\Omega_{1+}|^2} {|\Omega_{2+}|^2|\Omega_{1+}|^2+
|\Omega_{2-}|^2|\Omega_{1-}|^2+|\Omega_{2+}|^2|\Omega_{1-}|^2}~.
\end{equation}
Using Eq.~(\ref{e-prop}) we derive equations of motion for
the fields.  For example,
\begin{eqnarray} \label{propO2-}
&&\frac{\partial \Omega_{2- }}{\partial z} =  2i \kappa
\delta \Omega_{2-} \frac{\wp_{a-b0}^2}{\wp^2}
\\* \nonumber && \frac{2 |\Omega_{1+}|^2
|\Omega_{1-}|^2 |\Omega_{2+}|^2 + |\Omega_{2+}|^2
|\Omega_{1-}|^4}{(|\Omega_{1-}|^2|\Omega_{2+}|^2 +
|\Omega_{1+}|^2|\Omega_{2+}|^2 +
|\Omega_{1-}|^2|\Omega_{2-}|^2)^2}
\end{eqnarray}
where $\kappa$ is the coupling constant with respect to the
transition as a whole (i.e., $\gamma_r$ in Eq.~(\ref{kappa})
is now the total natural decay rate of the excited state),
and $\wp = (4\nu^3 \gamma_r/(3\hbar c^3))^{1/2}$ is the dipole
moment of the transition.

	The calculations can be considerably simplified if the
numerical values of the transition probabilities are used.
Let us now consider the particular case of the $M$ part
of the $F=2 \rightarrow F=1$ transition.  According to the
transition probabilities, shown in Fig.~\ref{levels21.fig}c,
we get $|\Omega_{2+}|^2/|\Omega_{2-}|^2=6|E_{+}|^2/|E_{-}|^2$
and $|\Omega_{1+}|^2/|\Omega_{1-}|^2=|E_{+}|^2/6|E_{-}|^2$.
The interaction Hamiltonian ($H_{M}\simeq \hbar
\tilde{\lambda}_{M}$) for the elliptically polarized laser
field can therefore be rewritten as
\begin{equation}
H_{M} \simeq 2\hbar \delta \frac{|E_{-}|^4-|E_{+}|^4}
{|E_{+}|^4+|E_{-}|^4 + 6|E_{+}|^2|E_{-}|^2} \label{eigen-M1a}
\end{equation}
and therefore
\begin{eqnarray} \nonumber
 \frac{\partial E_\pm}{\partial z}& = &\mp 8 i \pi \hbar \delta N \frac
{\nu}{c} E_\pm \times \\  && |E_{\mp}|^2\frac{3(|E_+|^4 +|E_-|^4) + 2
|E_+|^2|E_-|^2}{(|E_{+}|^4+|E_{-}|^4 +
6|E_{+}|^2|E_{-}|^2)^2} \label{e1}\,.
\end{eqnarray}
In what follows we derive the same equation
using the more rigorous modified Schr\"odinger
formalism~\cite{fleischhauer99lanl}.

\subsection{Solution based on the modified Schr\"odinger equations}

	The interaction described above of the four
electromagnetic fields with the $M$ energy level configuration
may be also studied using Schr\"odinger equations.  This
approach enables us to find exact expressions for all the
atomic observables when we can ignore spontaneous emission.
The state vector of the atom can be written as:
\begin{eqnarray}
|\Psi\rangle &=&  a_+ e^{-i \nu t} |a_+\rangle + a_- e^{-i
\nu t} |a_-\rangle + \\ \nonumber && b_0 |b_0\rangle + b_+
|b_+\rangle + b_- |b_-\rangle\,.
\end{eqnarray}
Solving the Schr\"odinger equation
$$
|\dot \Psi\rangle = - \frac{i}{\hbar} \hat H |\Psi\rangle
$$
for the interaction Hamiltonian Eq.~(\ref{ham-M1}), we obtain
the following equations of motion for the slowly-varying
state amplitudes:
\begin{eqnarray}
\dot a_+ &=& i\Omega_{1+} b_0 +i \Omega_{1-} b_+\label{wfe1}\\
\dot a_- &=& i\Omega_{2+} b_- +i \Omega_{2-} b_0\label{wfe11}\\
\dot b_+ &=& 2i\delta b_+ + i\Omega_{1-}^* a_+ \label{bp}\\
\dot b_- &=& -2i\delta b_- + i\Omega_{2+}^* a_- \label{bm}\\
\dot b_0 &=& i\Omega_{1+}^* a_+ + i \Omega_{2-}^* a_-~.
\label{wfe2}
\end{eqnarray}

	In the steady state regime, this system has a nontrivial
solution only for $\delta = 0$.  The solutions for nonzero
detunings correspond to zero amplitudes for all parameters.
Thus, to sustain steady state in the open system, external
pumping is necessary.  For a small splitting between ground
state levels $\hbar \delta \ll kT$, where $T$ is the temperature
of the vapor, we assume that in thermal equilibrium, i.e., in
the absence of all fields, all lower states $|b_\pm \rangle $
and $|b_0\rangle $ are equally populated.  And, therefore,
within the open-system approach, we assume that the atoms are
pumped into states $|b_+\rangle $, $|b_-\rangle $, or $|b_0
\rangle $ with equal probability from outside of the system.
The corresponding rate can be determined by the requirement
that the total probability to find an atom in any of the states
is unity.

	Unlike the density matrix approach, a straightforward
introduction of incoherent pumping into the ground
states of the system is impossible.  It was shown by
Fleischhauer~\cite{fleischhauer99lanl} in an elegant way that
the effective density matrix equations for open systems with
injection rates into states and decays out of states can be
written in terms of stochastic complex state amplitudes.

	Let us consider an effective density matrix equation
for an atomic ensemble undergoing a unitary interaction with
some external fields or potentials.  In addition, decay out
of atomic states $|j\rangle$ is taken into account with
rates $\gamma_j$.  Also injection into certain states is
considered with injection rates $R_{ij}$.  In our case the
injection occurs only into energy eigenstates of the atoms
or incoherent mixtures of them, so only diagonal elements of
the matrix $R_{ij}$ are nonzero.  If injection in the coherent
superposition states is considered, non-diagonal elements are
also required to be taken into account.

	An effective density matrix equation has the following
structure:
\begin{equation}
{\dot\rho}_{ij}(t) = R_{ij} -\frac{\gamma_i+\gamma_j}{2} \rho_{jj}
- \frac{i}{\hbar}\Bigl[ H,\rho\Bigr]_{ij}\label{DM1}
\end{equation}
where $\gamma_i$ are decay rates out of the system, which can
in general be different for individual states.   Generally,
the pump rates $R_{ij}$ are time dependent, but for the sake of
simplicity we assume in the following that the rates $R_{ij}$
are constant.

	Density matrix elements may be represented in terms
of state amplitudes $\rho_{ji}=c_i^* c_j$\,.  In order to
put the pump term $R_{ij}$ in a similar form, we introduce a
formal Gaussian stochastic variable $r_i$ with the following
properties:
\begin{eqnarray}
\langle r_i\rangle  &=& 0\label{G1}\\
\langle r_i r_j\rangle &=& 0\label{G2}\\
\langle r_i^* r_j\rangle &=& R_{ij}~.\label{G3}
\end{eqnarray}
This yields a set of amplitude equations with stochastic
pump terms:
\begin{eqnarray}
{\dot c}_i = r_i -\frac{\gamma_i}{2}\, c_i +\frac{i}{\hbar}
H_{ij}\,  c_j ~.\label{de}
\end{eqnarray}
Since the amplitude equations are linear, their solution will
be a linear functional of the stochastic pump rates $r_i$.
Thus the averaging of bilinear quantities such as $c_i^* c_j$
required to obtain the density matrix elements can easily
be performed.  Generally, solution $c_j$ of Eq.~(\ref{de})
no longer makes sense as the amplitude for the atomic wave
function.  It only determines density matrix elements of
the system.

	To apply the above technique to our problem, we rewrite
Eqs.~(\ref{wfe1})--(\ref{wfe2}) (with time derivatives set
equal to zero)
\begin{eqnarray}
 i\Omega_{1+}
 b_0 +i \Omega_{1-} b_+ = 0 \label{vfe1}\\
i\Omega_{2+} b_- +i \Omega_{2-} b_0 = 0\label{vfe11}\\
2i\delta b_+ + i\Omega_{1-}^* a_+ = ir_+ \label{vbp}\\
-2i\delta b_- + i\Omega_{2+}^* a_- = ir_- \label{vbm}\\
i\Omega_{1+}^* a_+ + i \Omega_{2-}^* a_- = ir_0 \label{vfe2}
\end{eqnarray}
where  the stochastic ``pumping'' is introduced
\begin{eqnarray}
\langle r_\pm\rangle &=& \langle r_0\rangle = 0 \nonumber\\*
\langle r_\pm r_\mp\rangle &=& \langle r_\pm r_0\rangle = 0 \nonumber\\*
\langle r_\pm^* r_\mp\rangle &=& \langle r_\pm^* r_0\rangle = 0 \nonumber\\*
\langle r_\pm^* r_\pm\rangle &=& \langle r_0^* r_0\rangle = r^2 ~.\nonumber
\end{eqnarray}

	Solving Eqs.~(\ref{vfe1})--(\ref{vfe2}) with respect
to $a_1$, $a_2$, $b_\pm$, and $b_0$ we get
\begin{eqnarray}
b_+ &=&-b_0 \frac{\Omega_{1+}}{\Omega_{1-}}~, \;\;\;
	b_- = -b_0 \frac{\Omega_{2-}}{\Omega_{2+}}
		\nonumber\\*
b_0 &=& \frac{r_+|\Omega_{2+}|^2\Omega_{1-}\Omega_{1+}^* + r_-
	|\Omega_{1-}|^2\Omega_{2+}\Omega_{2-}^* -
	r_0|\Omega_{1-}|^2|\Omega_{2+}|^2 }{2\delta \left (
	|\Omega_{1+}|^2|\Omega_{2+}|^2 -
	|\Omega_{1-}|^2|\Omega_{2-}|^2\right )}
		\nonumber\\*
a_- &=& \frac{r_+\Omega_{1+}^*\Omega_{1-}\Omega_{2-} +
	r_- |\Omega_{1+}|^2\Omega_{2+} - r_0|\Omega_{1-}|^2\Omega_{2-} }{
	|\Omega_{1+}|^2|\Omega_{2+}|^2 - |\Omega_{1-}|^2|\Omega_{2-}|^2 }
		\nonumber\\*
a_+ &=& -\frac{r_+|\Omega_{2-}|^2\Omega_{1-} + r_-
	\Omega_{2-}^*\Omega_{1+}\Omega_{2+} -r_0|\Omega_{2+}|^2\Omega_{1+}}
	{ |\Omega_{1+}|^2|\Omega_{2+}|^2 - |\Omega_{1-}|^2|\Omega_{2-}|^2}~.
\end{eqnarray}
Utilizing the normalization condition
\begin{equation} \label{norm}
\langle a_-^*a_- \rangle + \langle a_+^*a_+ \rangle +\langle
b_+^*b_+ \rangle+\langle b_-^*b_- \rangle + \langle b_0^*b_0
\rangle=1
\end{equation}
we get 
\begin{eqnarray} \label{r}
&&r =2\delta \left ( |\Omega_{1+}|^2|\Omega_{2+}|^2 -
|\Omega_{1-}|^2|\Omega_{2-}|^2\right )/ \\ && \nonumber \left \{
(|\Omega_{1-}|^2|\Omega_{2+}|^2 + |\Omega_{1+}|^2|\Omega_{2+}|^2 +
|\Omega_{1-}|^2|\Omega_{2-}|^2)^2+ \right. \\ && \nonumber \left.
4\delta^2 \left [
|\Omega_{1+}|^2|\Omega_{1-}|^2(|\Omega_{1+}|^2+|\Omega_{2+}|^2)+
\right. \right. \\ \nonumber && \left. \left. 2
(|\Omega_{1+}|^4|\Omega_{2+}|^2+|\Omega_{1-}|^4|\Omega_{2-}|^2)
\right ] \right \}^{1/2}~.
\end{eqnarray}
Using Eq.~(\ref{r}) we arrive at the complete solution of
the problem which takes into account all orders in $\delta$.
For $\delta=0$ the system is in a dark state and the density
matrix elements correspond to the elements generated by
Eq.~(\ref{M-dark}).  For a nonzero small two-photon detuning
the populations and coherences for the ground state stay
approximately unchanged.  The solution for the populations of
the excited states are
\begin{widetext}
\begin{eqnarray}
\rho_{a-a-} =
4\delta^2\frac{|\Omega_{1+}|^2|\Omega_{1-}|^2|\Omega_{2-}|^2 +
|\Omega_{1+}|^4|\Omega_{2+}|^2 +
|\Omega_{1-}|^4|\Omega_{2-}|^2}{(|\Omega_{1-}|^2|\Omega_{2+}|^2 +
|\Omega_{1+}|^2|\Omega_{2+}|^2 +
|\Omega_{1-}|^2|\Omega_{2-}|^2)^2} \\*
\rho_{a+a+} =
4\delta^2\frac{|\Omega_{2+}|^2|\Omega_{2-}|^2|\Omega_{1+}|^2 +
|\Omega_{2+}|^4|\Omega_{1+}|^2 +
|\Omega_{2-}|^4|\Omega_{1-}|^2}{(|\Omega_{1-}|^2|\Omega_{2+}|^2 +
|\Omega_{1+}|^2|\Omega_{2+}|^2 +
|\Omega_{1-}|^2|\Omega_{2-}|^2)^2}
\end{eqnarray}
\end{widetext}
and for the atomic polarizations are
\begin{widetext}
\begin{eqnarray}
\rho_{a-b0} = \frac{2\delta\Omega_{2-}(2 |\Omega_{1+}|^2
|\Omega_{1-}|^2 |\Omega_{2+}|^2 + |\Omega_{2+}|^2
|\Omega_{1-}|^4)}{(|\Omega_{1-}|^2|\Omega_{2+}|^2 +
|\Omega_{1+}|^2|\Omega_{2+}|^2 +
|\Omega_{1-}|^2|\Omega_{2-}|^2)^2} \label{pol1} \\*
\rho_{a+b0} = -\frac{2\delta\Omega_{1+} (2|\Omega_{1-}|^2
|\Omega_{2+}|^2 |\Omega_{2-}|^2 + |\Omega_{1-}|^2
|\Omega_{2+}|^4)}{(|\Omega_{1-}|^2|\Omega_{2+}|^2 +
|\Omega_{1+}|^2|\Omega_{2+}|^2 +
|\Omega_{1-}|^2|\Omega_{2-}|^2)^2} \label{pol2} \\*
\rho_{a-b-} = -\frac{2\delta\Omega_{2+}(2 |\Omega_{1+}|^2
|\Omega_{1-}|^2 |\Omega_{2-}|^2 + |\Omega_{2-}|^2
|\Omega_{1-}|^4)}{(|\Omega_{1-}|^2|\Omega_{2+}|^2 +
|\Omega_{1+}|^2|\Omega_{2+}|^2 +
|\Omega_{1-}|^2|\Omega_{2-}|^2)^2} \label{pol3} \\*
\rho_{a+b+} = \frac{2\delta\Omega_{1-}(2|\Omega_{1+}|^2
|\Omega_{2+}|^2 |\Omega_{2-}|^2 + |\Omega_{1+}|^2
|\Omega_{2+}|^4)}{(|\Omega_{1-}|^2|\Omega_{2+}|^2 +
|\Omega_{1+}|^2|\Omega_{2+}|^2 +
|\Omega_{1-}|^2|\Omega_{2-}|^2)^2} \label{pol4}~.
\end{eqnarray}
\end{widetext}
Here we kept only the lowest order terms in $\delta$.
In the expressions for the atomic polarizations, the first
term, containing the amplitude of all four optical fields
(for example, $\Omega_{1+}|\Omega_{1-}|^2 |\Omega_{2+}|^2
|\Omega_{2-}|^2$ in the equation for $\rho_{a+b0}$), is due
to the four-photon coherence (hexadecapole moment), whereas
the second term represents the effect of optical pumping.

	The propagation equation for the fields is
\begin{eqnarray}
\frac{\partial \Omega_{ij}}{\partial z} &=& i\frac {2 \pi \nu }{c} N
\frac{\wp_{ij}^2}{\hbar} \rho_{ij}
\end{eqnarray}
where the indexes $ij$ show that the values are related to the
same transition $|i\rangle \rightarrow |j\rangle$.  It is easy
to see, for example, that the matrix element in Eq.~(\ref{pol1})
results in the propagation equation in Eq.~(\ref{propO2-}).
The two approaches are therefore equivalent.  The equation of
motion for the circularly polarized electromagnetic fields in
$E_\pm$ are given by the following expressions:
\begin{eqnarray}
\frac{\partial E_+}{\partial z} &=& i\frac {2 \pi \nu }{c} N
\left( \wp_{a-b-} \rho_{a-b-} + \wp_{a+b0} \rho_{a+b0} \right )
\label{emM1} \\
\frac{\partial E_-}{\partial z} &=& i\frac {2 \pi \nu }{c} N \left( \wp_{a+b+}
\rho_{a+b+} + \wp_{a-b0} \rho_{a-b0} \right )\,.
\label{emM2}
\end{eqnarray}

	Substituting the expressions for atomic polarizations
Eqs.~(\ref{pol1})--(\ref{pol4}) and using the proper dipole
moments for each transition (for the ${}^{87}$Rb they are
equal $1/2$ for $|b_\pm\rangle \rightarrow |a_\pm \rangle$,
and $1/12$ for $|b_0\rangle \rightarrow |a_\pm \rangle$
(Fig.~\ref{levels21.fig}c)), we obtain Eqs.~(\ref{e1}).

	So far we have made no assumption concerning the losses
in the system.  Generally, this requires solving the Bloch
equations for the atomic populations and polarizations as was
done for the $\Lambda$ system.  For the $M$ scheme, however,
this process is rather involved even for the degenerate system
($\delta=0$).  Since the dark state exists for any value of Rabi
frequency $\Omega_{ij}$, it is always possible to transform
the basis of the atomic states so that there is one atomic
level uncoupled from the laser field.  The $M$ system can be
represented as two independent open two-level systems, connected
only via relaxation processes~\cite{morris'83}.  The absorption
in this systems has similar properties compared to those of the
$\Lambda$ system: it is proportional to decay rate $\gamma_0$
and inversely proportional to the light intensity.  The exact
analytical expression for this absorption is rather lengthy
and we do not present it here.

\subsection{Polarization rotation for an $F=2 \rightarrow F'=1$
transition}

	To describe the polarization rotation on the
$F=2 \rightarrow F'=1$ transition we write the interaction
Hamiltonian as a balanced sum of the Hamiltonians for the $M$
and $\Lambda$ systems, taking into account the branching ratio
for the atomic transitions
\begin{equation} \label{ham1}
H_{2 \rightarrow 1} = \zeta_1 H_\Lambda + \zeta_2 H_M = \zeta_1
\hbar \tilde{\lambda}_\Lambda + \zeta_2 \hbar \tilde{\lambda}_M
\end{equation}
where $\zeta_1$ and $\zeta_2$ are weighting coefficients
($\zeta_1+\zeta_2=1$) that describe the population
redistribution between the $\Lambda$ and $M$ schemes.
Using the numerical simulation of this system, we find them
to be equal with very good accuracy.  Using Eq.~(\ref{e-prop})
we now derive the equation of motion for this system:
\begin{widetext}
\begin{equation} \label{eqn_mot1}
\frac{\partial E_\pm}{\partial z} = \mp 4 i \pi \hbar \delta N \frac
{\nu}{c} E_\pm \frac{|E_\mp|^2}{(|E_{+}|^2+|E_{-}|^2)^2}
\left[ 1+ 2(|E_{+}|^2 + |E_{-}|^2)^2\frac{3(|E_+|^4 +|E_-|^4) + 2
|E_+|^2|E_-|^2}{(|E_{+}|^4+|E_{-}|^4 +
6|E_{+}|^2|E_{-}|^2)^2} \right].
\end{equation}
\end{widetext}
It is interesting to note that for linearly polarized light
($|\Omega_+| = |\Omega_-|$) the contributions from $\Lambda$
and $M$ system are identical, and Eq.~(\ref{eqn_mot1}) coincides
with Eq.~(\ref{e1a}).  This proves that a single $\Lambda$
system may be used for accurate description of the dispersive
properties of more complicated level configurations.

	Let us introduce the electromagnetic field ellipticity
parameter $q$ such that the amplitudes of the circularly
polarized components are $E_{\pm} = |E|\sqrt{(1\pm q)}
\exp (i\phi _{\pm})/ \sqrt 2$.  Then Eq.~(\ref{eqn_mot1})
transforms to
\begin{equation}
\frac{\partial E_\pm}{\partial z} =
\mp 2 i \pi \hbar \delta N \frac {\nu}{c}
 \frac{E_\pm(1\mp q)}{|E|^2} \left [ 1 + 2 \frac{2+q^2}
{(2-q)^2}  \right ]~. \label{eqn_mot2}
\end{equation}
Based of the results of our numerical simulation, we
conclude that absorption of light that interacts with the $F=2
\rightarrow F'=1$ transition does not depend on the ellipticity
of the light.  The light transmission through the cell can be
described by an equation similar to Eq.~(\ref{lambda-transm}):
\begin{eqnarray} \label{M-absorption}
I_{\mathrm{\mathrm{out}}} &=&
  I_{\mathrm{in}}\left(1-\frac{2\pi \hbar \gamma_0 N L}{|E(0)|^2}
\frac {\nu}{c}
\right).
\end{eqnarray}
The rotation angle for the light polarization is then given by
\begin{eqnarray} \label{M-rotation}
 \phi &=& \frac{\delta}{\gamma_0}\left[1+ 2 \frac{2+q^2}{(2-q^2)^2}
\right ] {\rm ln} \frac{I_{\mathrm{in}}}{I_{\mathrm{out}}}
\end{eqnarray}
where $I_{\mathrm{in}}$ and $I_{\mathrm{out}}$ are
the intensities of the electromagnetic field at the entrance
and exit of the medium.  The value of polarization rotation
increases with the light ellipticity by the factor
\begin{equation} \label{ratio}
\frac{\phi_{M+\Lambda}}{\phi_{\Lambda}}=\frac12 \left(1+ 2
\frac{2+q^2}{(2-q^2)^2} \right )
\end{equation}
compared to $\Lambda$ system.  Therefore NMOR on the $F=2
\rightarrow F'=1$ transition may only be properly described
by a $\Lambda$ configuration for linearly polarized light.
The difference between the $M$ and $\Lambda$ systems results
from the hexadecapole moment induced in $M$ configuration.

\section{NMOR in atoms with large values of angular momentum}

\begin{figure}
\centering
\includegraphics[width=1.00\columnwidth]{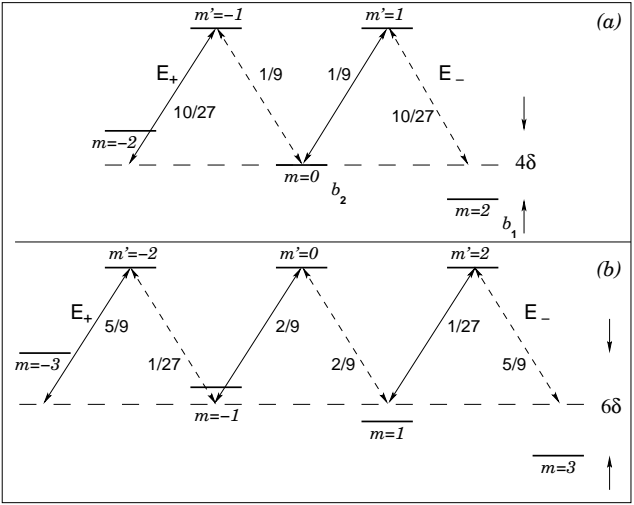}
\caption{\label{levels32.fig}
	Energy level scheme for $^{85}$Rb atoms.  This
	scheme may be decomposed into a superposition of a)
	$M$-system and b) triple-$\Lambda$ system.  Transition
	probabilities are shown for each individual transition.
}
\end{figure}

\begin{figure}
\centering
\includegraphics[width=1.00\columnwidth]{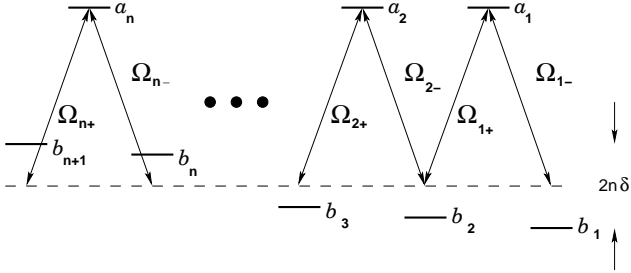}
\caption{\label{multiLambda.fig}
	Generalized $M$ interaction scheme.  Here $\Omega_{i-}
	= E_- \wp_{a_ib_i}/\hbar$, $\Omega_{i+} = E_+
	\wp_{a_ib_{i-1}}/\hbar$.
}
\end{figure}

	Higher order coherence can be excited among
Zeeman sublevels of alkali atoms with $F > 2$.  Here we
find a perturbed dark state for the generalized $M$ scheme
consisting of an arbitrary number of $\Lambda$-links, using
the method described above.  Then we apply these results to
evaluate the nonlinear Faraday rotation in the ${}^{85}$Rb
$F=3 \rightarrow F=2$ transition.  We consider the scheme in
Fig.~\ref{multiLambda.fig}.  The interaction Hamiltonian for
this scheme is
\begin{eqnarray}
H_{n \times \Lambda} &=& -\hbar \delta \sum_{k=0}^n(n-2k)
|b_{k+1}\rangle \langle b_{k+1}|
\\*
&+& \hbar \sum_{k=1}^n \left(\Omega_{k-}|a_k\rangle \langle b_k| +
\Omega_{k+}|a_k\rangle \langle b_{k+1}| + H.c. \right) \nonumber
\label{ham-nLambda}
\end{eqnarray}
Here $n$ is the number of $\Lambda$ links, which connects $n+1$
ground-state levels via $n$ excited states.  There exists a
dark state for this system for exact resonance ($\delta$=0):
\begin{equation}
|D\rangle = \frac{ \displaystyle
\sum \limits_{k=0}^n(-1)^k \displaystyle
\prod_{j=1}^k \Omega_{j-}  \displaystyle  \prod \limits_{l=k+1}^n
\Omega_{l+}\ |b_{k+1}\rangle}
{\sqrt{\displaystyle \sum\limits _{k=0}^n \displaystyle \prod_{j=1}^k
|\Omega_{j-}|^2  \displaystyle  \prod\limits_{l=k+1}^n
|\Omega_{l+}|^2}}
\end{equation}
where we use a convention that $\prod_{j=1}^0 \equiv
\prod_{j=n+1}^n\equiv 1$.  We deduce the perturbed ``dark
state'' eigenvalue for the Hamiltonian using the same procedure
as we used before in Eq.~(\ref{ham-nLambda})
\begin{equation}
\tilde{\lambda}_{n\times \Lambda}\simeq \delta \frac{
\displaystyle \sum \limits_{k=0}^n(2k-n) \displaystyle \prod
\limits_{i=1}^k|\Omega_{i-}|^2 \displaystyle \prod
\limits_{j=k+1}^n|\Omega_{j+}|^2} {\displaystyle \sum
\limits_{k=0}^n \displaystyle \prod \limits_{i=1}^k|\Omega_{i-}|^2
\displaystyle \prod \limits_{j=k+1}^n|\Omega_{j+}|^2}~.
\end{equation}
The equation of motion for the circularly polarized
electromagnetic fields can be found from Eq.~(\ref{e-prop}).
As an example, let us calculate the interaction Hamiltonian for
light interacting with the $5S_{1/2} F=3 \rightarrow 5P_{1/2}
F'=2$ transition of ${}^{85}$Rb (Fig.~\ref{levels32.fig}).  The
circularly polarized components of the resonant electromagnetic
field form an $M$ scheme and a triple-$\Lambda$ scheme. Using
the proper values of the transition probabilities, shown in
the same Figure, we derive
%
%
\begin{eqnarray} \label{ham_triple}
&& H_{3 \rightarrow 2} = 3\hbar \delta \left[ 2 \zeta_1
\frac{|E_-|^4 - |E_+|^4} {3|E_+|^4 + 3|E_-|^4 + 10
|E_+|^2 |E_-|^2} + \right. \nonumber \\*
&& \left. \zeta_2 \frac{ |E_-|^6 + 5|E_+|^2|E_-|^4 -
5|E_+|^4|E_-|^2 - |E_+|^6}{|E_+|^6 + 15|E_-|^2|E_+|^4 +
15|E_-|^4|E_+|^2 + |E_-|^6} \right] ~.\nonumber \\*
&&
\end{eqnarray}
Here again $\zeta_1,2$ are the coefficients reflecting the
population distribution between to schemes.  By differentiating
the Hamiltonian it is easy to find the polarization rotation
in the system
\begin{equation}
\frac{\partial \phi}{\partial z} = - 6i\pi  N\frac{\nu}{c}
\frac{\hbar \delta}{|E|^2}
\left [ 2\zeta_1 \frac{4+q^2}{(4-q^2)^2} + \zeta_2
\frac{8-6q^2+3q^4} {(4-3q^2)^2}  \right]. \label{rot-prop-mL}
\end{equation}
It is obvious that both interaction chains contribute to
the elliptically dependent NMOR.  At the same time different
orders of the nonlinear susceptibility are responsible for
the polarization rotation:  if in the case of the $M$ scheme
it is $\chi^{(3)}$ nonlinearity, for the triple-$\Lambda$
scheme it is $\chi^{(5)}$ nonlinearity, since there are $7$
photons involved in the creation of the ground-state coherence.
This might be the reason why the triple-$\Lambda$ scheme
shows more enhancement of the polarization rotation for
nearly circular polarization compared to the rotation of
linear polarization than does the $M$ scheme ($10$ vs $20/9$
times for the $F=3 \rightarrow F'=2$ transition).

\section{Application for quantum information processing}

	So far we have considered $\Lambda$ and $M$
schemes of the type described in Figs.~\ref{levels21.fig}a
and~\ref{levels21.fig}c.  Here the two-photon detunings with
respect to states $|b_{+} \rangle $ and $|b_{-} \rangle$
are equal and opposite in sign and all the fields are treated
classically.  This approach is useful for describing NMOR in
alkali atomic vapors.  In general, however, the $M$ system
may be created by strongly nondegenerate atomic levels and
all four fields connecting corresponding atomic transitions
may be independent.  This case is especially interesting if
we are going to use the enhanced Kerr nonlinearity the system
provides~\cite{matsko_prep'02}.

	In this section we compare the $N$ and $M$
configurations shown in Figs.~\ref{fig1.fig}b and
\ref{fig1.fig}c. The $N$ system is essentially a $\Lambda
$ system with an additional nonresonant transition.
Similarly, the $M$ system in Fig.~\ref{fig1.fig}c is a
resonant $N$ system with an additional detuned transition.
Since these systems have potential applications in the
field of quantum information processing, we discuss them
here. Some details concerning such systems have been given
earlier~\cite{zubairy02pra,matsko_prep'02}.  The systems
seems to be completely different because the all-resonant $N$
configuration demonstrates enhanced three-photon absorption,
while the all resonant $M$ configuration demonstrates complete
transparency.  We show here that the performance of these
schemes as sources of refractive Kerr nonlinearity is very
similar.

	We assume that the $| a_1 \rangle \rightarrow |b_1
\rangle$ and $ | a_2 \rangle \rightarrow |b_2 \rangle
$ transitions are induced by quantized fields whereas
the transitions $| a_1 \rangle \rightarrow |b_2 \rangle$
and $|a_2 \rangle \rightarrow |b_3 \rangle $ are induced by
classical fields of Rabi frequencies $\Omega_1$ and $\Omega_2$,
respectively.

	The Hamiltonians for the $N$ and $M$ schemes in the
slowly varying amplitude and phase approximations are
\begin{eqnarray}
H_N &=& \hbar \Delta |a_2\rangle \langle a_2| + \hbar
(\hat{\alpha}_1 |a_1\rangle \langle b_1| \nonumber \\
&+&  \Omega_1|a_1\rangle \langle b_2| +
\hat{\alpha}_2|a_2\rangle \langle b_2| + H.c. )
\label{ham-N}  \\
H_M &=& -\hbar \delta |b_3\rangle \langle b_3| + \hbar (
\hat{\alpha}_1|a_1\rangle \langle b_1| + \Omega_{1}|a_1\rangle
\langle b_2| \nonumber \\
 &+& \hat{\alpha}_{2}|a_2\rangle \langle b_2| +
\Omega_{2}|a_2\rangle \langle b_3| + H.c. ) \label{ham-M2}
\end{eqnarray}
where $H.c.$ means Hermitian conjugate and the relation between
Rabi frequencies of the probe fields and quantum operators
describing the corresponding field mode can be written as
\begin{equation}
\hat \alpha_i = \sqrt{\frac{2\pi \wp^2_i \nu_i}{\hbar V_i}} \hat
a_i = \eta_i \hat a_i
\end{equation}
where $\wp_i$ is the dipole moment of the transition
$|a_i\rangle \rightarrow |b_i\rangle$, $\nu_i$ is the field
frequency, $V_i$ is the quantization volume of the mode, $\hat
a_i$ and $\hat a_i^\dag$ are the annihilation and creation
operators.  Proceeding along the same lines as in Sec.~III we
obtain the effective Hamiltonian for the two configurations
of the form
\begin{equation} \label{heff}
H_{\mathrm{eff}}= \hbar \tilde{\delta} \hat{a}^{\dagger}_1 \hat{a}_1
\hat{a}^{\dagger}_2 \hat{a}_2
\end{equation}
where the the coupling constant $\tilde{\delta}$ for the two
configurations is of the form~\cite{zubairy02pra}:
\begin{equation}
\tilde{\delta}_N= \frac{\eta_1^2}{\Delta}\frac{\eta_2^2}{\Omega_1^2}
\end{equation}
and~\cite{matsko_prep'02}
\begin{equation}
\tilde{\delta}_M= -\delta \frac{\eta_1^2}{|\Omega_1|^2}
\frac{\eta_2^2}{|\Omega_2|^2}~.
\end{equation}

	Any system that may be described by the Hamiltonian in
Eq.~(\ref{heff}) has a potential application in implementing
a quantum phase gate.  The transformation for a two-bit
quantum phase gate for the $j$th and $k$th qubits is
given by $Q_{\eta}^{jk}|\alpha_j,\beta_k\rangle={\rm
exp}(i\eta \delta_{\alpha _j,1} \delta_{\beta
_k,1})|\alpha_j,\beta_k\rangle$, where $|\alpha_j\rangle$
and $|\beta_k\rangle$ stand for the basis states $|0\rangle$
or $|1\rangle$ of the qubits.  Thus the quantum phase gate
introduces a phase $\eta$ only when both the qubits in the
input states are $1$.  A representation of the quantum phase
gate is given by the operator
\begin{eqnarray}
Q_{\eta}^{jk}&=&|0_j,0_k\rangle \langle 0_j,0_k|+
|0_j,1_k\rangle \langle 0_j,1_k| \nonumber \\
&+&|1_j,0_k\rangle \langle 1_j,0_k|+e^{i \eta}|1_j,1_k\rangle
\langle 1_j,1_k|~.
\end{eqnarray}
It is clear that such a phase gate can be realized via
Hamiltonian $H_{\mathrm{eff}}$ with the time-evolution
unitary operator $exp(-iH_{\mathrm{eff}}\tau/\hbar)$ and the
corresponding phase $\eta= \tilde{\delta} \tau $ where $\tau$
is the interaction time.

	The nonlinearities associated with both the present
$N$ and $M$ schemes correspond to $\chi^{(3)}$.  The resonant
enhancement of $\chi^{(5)}$ and higher order nonlinearities
can be obtained by adding more $\Lambda$ sections to $N$
or $M$ schemes.  In general, the effective Hamiltonian for
$\chi^{(2m-1)}$ is
\begin{equation}
H^{(2m-1)}_{\mathrm{eff}}=\hbar \tilde{\delta}^m \hat{a}^{\dagger}_1
\hat{a}_1 \hat{a}^{\dagger}_2 \hat{a}_2...\hat{a}^{\dagger}_m
\hat{a}_m
\end{equation}
where, for extended $N$ systems,
\begin{equation}
\tilde{\delta}^{m}_N =
  (-1)^{m-1}\frac{\eta_1^2}{\Delta}\frac{\eta_2^2}{|\Omega_1|^2}
...\frac{\eta_m^2}{|\Omega_{m-1}|^2}
\end{equation}
and, for the extended $M$ system,
\begin{equation}
\tilde{\delta}^{m}_M=(-1)^{m} \delta \frac{\eta_1^2}{|\Omega_1|^2}
\frac{\eta_2^2}{|\Omega_2|^2}...\frac{\eta_m^2}{|\Omega_m|^2}~.
\end{equation}
Such nonlinearities can be used in implementing m-bit quantum
phase gates that are defined via
\begin{eqnarray}
&& Q^{(m)}_{\eta}|\alpha_1,\alpha_2,...\alpha_m \rangle = \nonumber\\*
&& ~~~~
\exp (i\eta \delta_{\alpha_1,1}\delta_{\alpha_2,1}...\delta_{\alpha_{m,1}})
|\alpha_1,\alpha_2,...\alpha_m \rangle~.
\end{eqnarray}
In other words, a phase $\eta $ is introduced when all the qubits are
in state $|1\rangle$.  Thus if qubit states $|0\rangle$ and
$|1\rangle$ are defined via photon number states, the m-bit
quantum phase gate is implemented via  $Q^{(m)}_{\eta } =
\exp(-i\hat H_m \tau/\hbar)$, $\eta = \tilde \delta \tau$.
Such gates may have important applications in quantum computing
algorithms such as those related to quantum search of unsorted
database~\cite{dzc}.

	The important question is how large can the phase shift
$\eta$ be.  Our initial estimates indicate that phase shifts as
large as 3 radians can be obtained for $m=3$ via $\chi^{(5)}$
nonlinearities.  However, there are problems related to phase
mismatch between different photons which arise because the group
velocities can be different for different pulses.  Such problems
can be overcome by methods discussed in~\cite{lukin00prl}.

	It is interesting to mention that interaction Hamiltonian 
for symmetrical $M$-scheme, given by Eq.(\ref{eigen-M1a}) is 
identical to one for asymmetric $M$ scheme, considered in this
 section - Eq.(\ref{heff}) in case when one of the circularly 
 polarized component is much stronger than the other (nearly 
 circularly polarized light). This means that the quantum phase gate
 discussed above, can be potentially created even using Zeeman 
 substructure of alkali atoms, resolved in magnetic field. Unfortunately, 
 in case of generalized $M$ scheme this is not true.

\section{Susceptibilities for inhomogeneously broadened
$\Lambda$, $N$ and $M$ systems}

	It is important to know what changes are introduced by
Doppler broadening to the systems discussed above.  For the
sake of simplicity we restrict ourselves to asymmetric
schemes discussed in the previous section.  Let us start
with the Doppler broadened $\Lambda$ system shown in
Fig.~\ref{fig1.fig}a.  This system is widely discussed in the
literature~\cite{lee_prep'02,javan_prep'01,budker_prep'02}, so
we consider only the necessary steps that allow us to calculate
susceptibility for the Doppler broadened $M$ configuration.
To sustain EIT in a Doppler-broadened $\Lambda$ medium the
minimum value of the Rabi frequency of the coupling field
$\Omega_1$ ($|\Omega_1| \gg |\alpha_1|$) should exceed $W_d
\sqrt{\gamma_0/\gamma}$, where $W_d$ is the linewidth of
the Doppler distribution ($W_d \sqrt{\gamma_0/\gamma} \gg
\sqrt{\gamma_0 \gamma}$)~\cite{javan_prep'01}.  Then the
population of the state $|b_1\rangle$ is almost unity and
density matrix element (c.f.\ Eq.~\ref{rhoabpm}) for the probe
transition reduces to
\begin{equation}
\rho_{ab1} \simeq
\frac{i \alpha_1 (\gamma_0-i\delta)}{(\gamma+i(\delta+kv))
(\gamma_0-i\delta)+|\Omega_1|^2}
\end{equation}
where $k$ is the wave vector of the field, and $v$ is the atomic
velocity.  We simplify the problem by using a Lorentzian profile
as the velocity distribution function $f(kv)$ with full width
at half maximum $2W_D$ such that $f(kv) = (1/\pi) W_D/[W_D^2 +
(kv)^2]$.  Integrating over the Doppler distribution we get
\begin{eqnarray}
\langle  \rho_{a\, b1} \rangle_v &=& \frac{i\alpha_1
(\gamma_0-i\delta)}{(\gamma+W_D-i\delta)
(\gamma_0-i\delta)+|\Omega_1|^2} \\ \nonumber
&\approx& \frac{i\alpha_1}{\gamma+W_D
+ i |\Omega_1|^2/\delta}~.
\end{eqnarray}
This result was evaluated using the contour integration
in the complex plane which contains one pole in the lower
half, $(kv)_1 = - i W_D$.  Let us consider the $M$ scheme
shown in Fig.~\ref{fig1.fig}c ($|\Omega_i| \gg |\alpha_j|$).
The susceptibility for the field $\alpha_2$ may be obtained
similarly to the $\Lambda$ scheme.  The population
of level $|b_2 \rangle$ is equal to, approximately,
$|\alpha_{1}|^2/|\Omega_{1}|^2$.

	The nonlinear interaction appears as the result
of the refraction and absorption of the second probe field
$\alpha_{2}$, coupled to the second drive field $\Omega_{2}$,
that create a $\Lambda$ system.  Therefore, we get the
susceptibility
\begin{equation} \label{sus2}
\chi_M = -i\frac{3}{8\pi^2} N\lambda_{\alpha 2}^3 \frac{\gamma_2
(\gamma_0-i\delta)}{(\gamma_0-i\delta)W_d+|\Omega_2|^2}
\frac{|\alpha_1|^2}{|\Omega_1|^2}
\end{equation}
where $N$ is the atomic density, $\gamma_2$ is the decay
rate of the level $|a_2 \rangle$, $\lambda_{\alpha 2}$ is the
vacuum wavelength of the field $\alpha_2$.

	Finally, let us consider the $N$ level configuration
shown in Fig.~\ref{fig1.fig}c.  If the condition $\Delta \gg
\gamma_2$ is fulfilled, then the population of level $|b_2
\rangle$ is equal to approximately $|\alpha_1|^2/|\Omega_1|^2$.
The nonlinear interaction appears as the result of the
refraction and absorption of the second probe field $\alpha_2$,
far detuned from the corresponding atomic transition.  For the
corresponding two-level system we derive
\begin{eqnarray}
\rho_{a2b2} \simeq \frac{i \alpha_2}{\gamma + i (\Delta + kv) }
\frac{|\alpha_1|^2}{|\Omega_1|^2}~.
\end{eqnarray}
The corresponding susceptibility for the field $\alpha_2$ is
\begin{equation} \label{sus1}
\chi_N = -i\frac{3}{8\pi^2}N\lambda_{\alpha 2}^3
\frac{\gamma_2}{W_d+i\Delta} \frac{|\alpha_1|^2}{|\Omega_1|^2}~.
\end{equation}
The nonlinear phase shift may be increased, formally, by
increasing the atomic density or interaction length.  This is
impossible to implement practically because of the absorption
of the medium.  Therefore, to compare the nonlinear performance
of different nonlinear systems one needs to compare the ratio
of their refractive nonlinearities and corresponding residual
absorption, linear as well as nonlinear.  The effective ratio
between absorption and nonlinearity for the Doppler broadened
$N$ scheme (\ref{sus1}) is $W_d/\Delta$.  It is easy to
see that (\ref{sus1}) and (\ref{sus2}) are interchangeble
if $\gamma_0 \rightarrow 0$, and $\Delta \leftrightarrow
\delta/|\Omega_2|^2$.  Therefore, the $M$ and $N$ schemes
are equivalent in sense of the effective Kerr nonlinearity
they produce.

\section{Experimental study of NMOR with elliptically polarized
light in Rb vapor}

\subsection{Experimental setup}

\begin{figure}
\centering
\includegraphics[width=1.00\columnwidth]{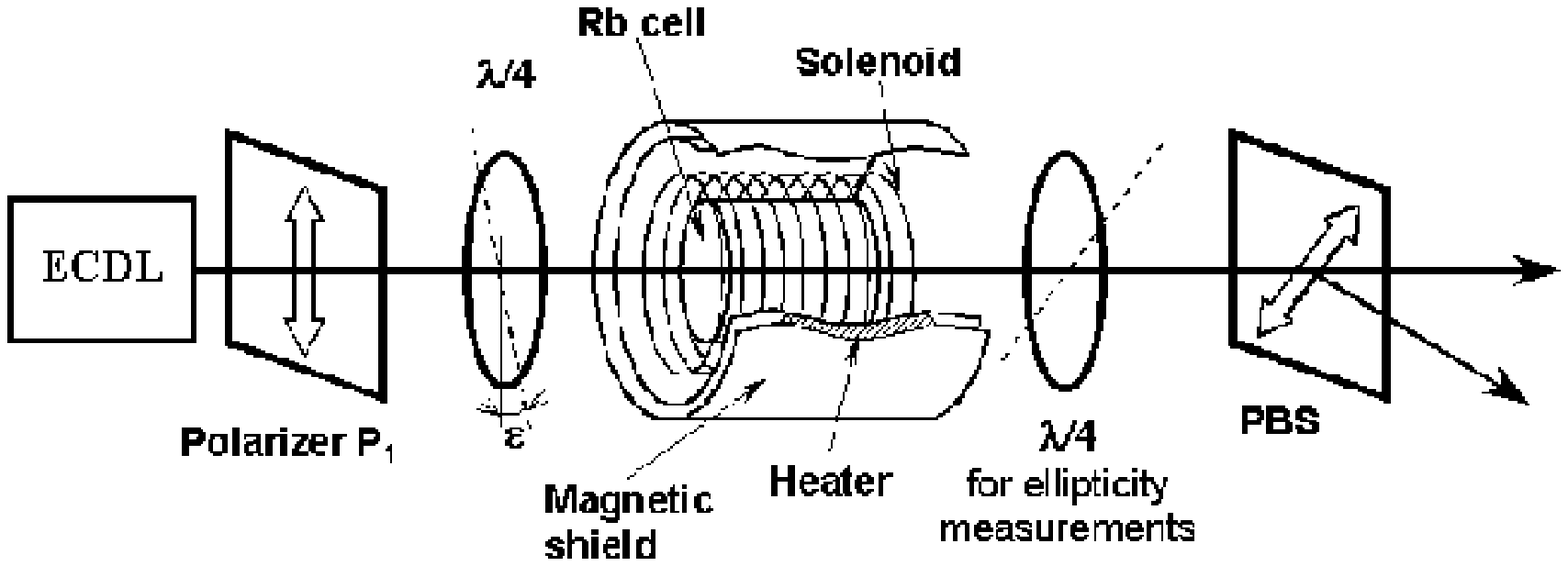}
\caption{\label{setup.fig}
	Schematic of the experimental setup.
}
\end{figure}

	The experimental setup is shown schematically in
Fig.~\ref{setup.fig}.  We use an external cavity diode
laser (ECDL) tuned in the vicinity of the $D_1$ line of
${}^{87}$Rb ($\lambda=795~\mathrm{nm}$).  The initial linear
polarization is produced by a high-quality polarizer $P_1$;
the initial ellipticity of the beam $\epsilon$~\cite{note}
is then controlled by a quarter wave-plate placed after
the polarizer.  Maximum laser power delivered to the atomic
cell is $P_{max}=2~\mathrm{mW}$.  A cylindrical glass cell of
length $50$~mm and diameter $25$~mm is filled with isotopically
enhanced ${}^{87}$Rb.  It is placed inside a two-layer magnetic
shield to minimize the influence of the laboratory magnetic
field.  The atomic density is controlled by a heating element
placed between the two shielding layers.  The longitudinal
magnetic field is created by a solenoid mounted inside the
inner magnetic shield.

	To measure the transmitted laser power and the
polarization rotation angle a polarization beam splitter (PBS)
is placed after the atomic cell.  The signals from the two
PBS channels $S_{1,2}$ are collected while the axis of the PBS
is tilted at $45^o$ degrees with respect to the main axis of
the initial polarization ellipse.  In this configuration the
transmitted light power is proportional to the sum of the two
signals $S_1+S_2$ and the polarization rotation angle $\phi$
is given by:
\begin{equation}\label{exp_rot}
\phi = \frac12 \arcsin \frac{S_1-S_2}{(S_1+S_2)\cos 2\epsilon}~.
\end{equation}
It is also possible to detect the ellipticity of the outgoing
laser beam by placing another quarter waveplate after the
cell and before the PBS.  When the fast waveplate axis is
aligned with the PBS axis and makes $45^o$ with the initial
polarization direction, the ellipticity $\epsilon$ of the beam
can be found similarly to the rotation angle:
\begin{equation}\label{exp_ell}
\epsilon = \frac12 \arcsin \frac{\tilde{S}_1- \tilde{S}_2}{
\tilde{S}_1+ \tilde{S}_2}~.
\end{equation}

\subsection{The experiments with ${}^{87}$Rb vapor}

	There are two factors contributing to the rotation
of the elliptical polarization of light propagating through
the Rb vapor:  the nonlinear Faraday rotation, caused by the
shifts of the magnetic sublevels in an external magnetic field,
and the self-rotation caused by the ac-Stark shifts due to the
off-resonant interaction of the electro-magnetic field with
far-detuned levels~\cite{boyd'92,budker'01,novikova'02a}.
Since the latter effect does not depend on the magnetic
field, we eliminate it from the experimental data either by
our measurement procedure or by direct subtraction.  In all
further discussions we concentrate on NMOR signals only.

	Let us first study the modification of the polarization
rotation by measuring the rotation rate $\frac{d\phi}{dB}(B=0)$
for different degree of light ellipticity.  We find the rotation
rate by modulating the magnetic field by a small amount and
dividing the difference of two rotation signals corresponding
to the small variation of the magnetic field by the magnitude
of this variation.  This way we detect only the rotation which
depends on the external magnetic field.

\begin{figure}
\centering
\includegraphics[width=1.00\columnwidth]{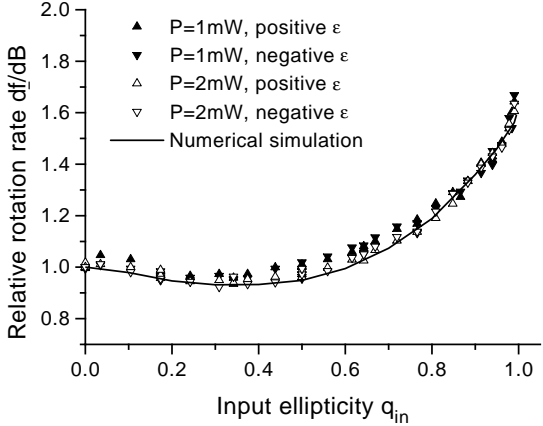}
\caption{\label{dfidB1.fig}
	The normalized slope of the nonlinear magneto-optic
	rotation as a function of the ellipticity of the
	incident light.  Experimental data are shown for
	opposite values of ellipticity and two different
	values of laser power: P=$2$~mW (solid up triangles
	for positive ellipticity and solid down triangles
	for negative ellipticity) and P=$1$~mW (hollow up
	triangles for positive ellipticity and hollow down
	triangles for negative ellipticity). The results of
	the numerical simulations for the case of $2$~mW laser
	power are shown by a solid line.  Absolute values
	of the nonlinear Faraday rotation for the linear
	polarization were $d\phi/dB(B=0)=4.5~\mathrm{rad/G}$
	and $6~\mathrm{rad/G}$ for P=$2$~mW and P=$1$~mW
	respectively.
}
\end{figure}

	The rotation rate as a function light ellipticity is
shown in Fig.~\ref{dfidB1.fig}.  We observe a polarization
rotation enhancement as predicted theoretically.  At the
same time, the experimental data cannot be fitted using
Eq.~(\ref{ratio}) because of the Doppler broadening of
the transition and the ac-Stark of the magnetic sublevels.
However, an exact numerical simulation based on steady state
solution of Maxwell-Bloch equations for the $F=2 \rightarrow
F'=1$ transition, which takes into account these effects,
is in excellent agreement with the experimental data.

\begin{figure}
\centering
\includegraphics[width=1.00\columnwidth]{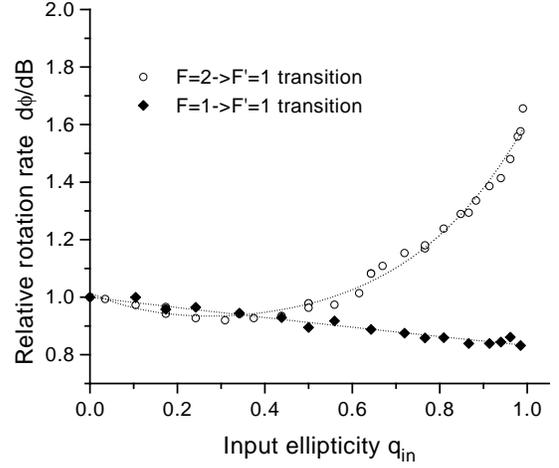}
\caption{\label{dfidB_tr.fig}
	The normalized slope of nonlinear magneto-optic
	rotation as a function of the ellipticity of the
	incident light for the $\Lambda$ scheme (transition
	$F=1 \rightarrow F'=1$) and $M+\Lambda$ scheme
	(transition $F=1 \rightarrow F'=1$).  Dotted lines are
	to guide the eyes.  Input laser power is P=$2$~mW,
	the atomic densities are chosen to provide $85\%$
	absorption on each transition.  The absolute
	value of the nonlinear Faraday rotation of linear
	polarization were $d\phi/dB(B=0)=1.8~\mathrm{rad/G}$
	and $4.5~\mathrm{rad/G}$ for the $F=1,2 \rightarrow
	F'=1$ transitions respectively.
}
\end{figure}

	It is also possible to verify that there is no
polarization rotation enhancement in an isolated $\Lambda$
system.  To do that we tune the laser to the $F=1 \rightarrow
F'=1$ transition of the ${}^{87}$Rb $D_1$ line.  In this case,
the ground-state coherence is formed by only one $\Lambda$
link.  The relative rotation rate for $F=1,2 \rightarrow
F'=1$ transitions are presented in Fig.~\ref{dfidB_tr.fig}.
Although there is a slight dependence of the rotation angle on
the light ellipticity for $F=1 \rightarrow F'=1$ transition,
this deterioration may be determined by Doppler broadening,
ac-Stark shifts, etc.

\begin{figure}
\centering
\includegraphics[width=1.00\columnwidth]{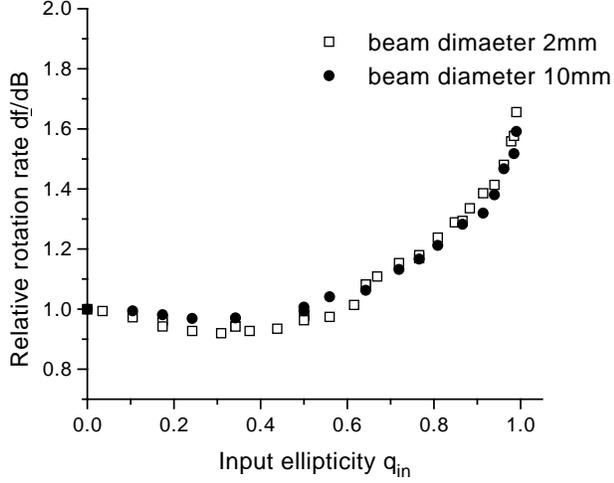}
\caption{\label{dfidB2.fig}
	The normalized slope of nonlinear magneto-optic rotation
	as a function of the ellipticity of the incident light
	for two different beam diameters: $d=2~\mathrm{mm}$
	(circles) and $d=10~\mathrm{mm}$ (diamonds).  In both
	cases the laser power is kept at $2$~mW.  Absolute
	values of the nonlinear Faraday rotation for the linear
	polarization were $d\phi/dB(B=0)=4.5~\mathrm{rad/G}$
	and $30~\mathrm{rad/G}$ respectively.
}
\end{figure}

	It is important to point out that even though
the theoretical expression for the relative rotation
rate (Eq.~(\ref{ratio})) does not fit the experimental
data precisely, it correctly predict some of the rotation
properties.  For example, our experiments confirm that the
relative rotation rate does not depend on the sign of the
ellipticity (Fig.~\ref{dfidB1.fig}).  If we vary the total
laser power or the coherence decay rate $\gamma_0$ (by varying
the laser beam diameter), the absolute value of the rotation
changes according to Eq.~(\ref{M-rotation}); its dependence
on the light ellipticity is the same within the experimental
uncertainty (Figs.~\ref{dfidB1.fig} and \ref{dfidB2.fig}).

\begin{figure}
\centering
\includegraphics[width=.70\columnwidth]{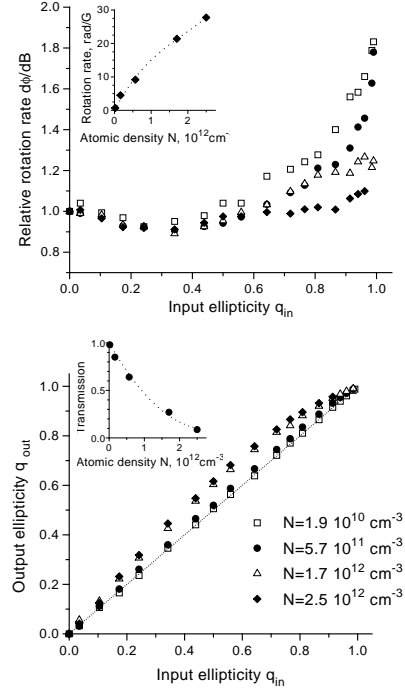}
\caption{\label{dfidB3.fig}
	(a) The normalized slope of nonlinear magneto-optic
	rotation as a function of the ellipticity of the
	incident light for various atomic densities.  Laser
	power is $2$~mW, beam diameter $d=2~\mathrm{mm}$.
	Inset: Absolute value of the nonlinear Faraday
	rotation of linear polarization as a function
	of atomic density. (b) The output ellipticity
	$\epsilon$ as a function of the ellipticity of
	the incident light for various atomic densities.
	Dotted line is for unchanged ellipticity.  Inset:
	Transmission $I_{\mathrm{out}}/I_{\mathrm{in}}$ of
	linear polarization as a function of atomic density.
}
\end{figure}

	All previous data were obtained for optically thin
Rb vapor (transmission $I_{\mathrm{out}}/I_{\mathrm{in}}
\simeq 0.85$).  The dependence of the relative rotation
rate on the ellipticity for higher atomic densities is
shown in Fig.~\ref{dfidB3.fig}.  It is easy to see that
for nearly circular polarization the rotation decreases
as atomic density is increased.  This may be caused by
optical pumping to the other ground state hyperfine levels,
as well as by the destruction of atomic coherence by radiation
trapping~\cite{molish_book,marsko'01prl}.

	The precise value of the output ellipticity of the laser
polarization is required for accurate polarization rotation
measurements (see Eq.~(\ref{exp_rot})).  The experimental
observations demonstrate that for optically thin media the
ellipticity of the light does not noticeably change due to
propagation effects if the magnetic field is small.  As the
atomic density increases, however, the ellipticity increases
(Fig.~\ref{dfidB3.fig}b).  Although this change is relatively
small ($< 15 \%$), the associated error in the calculated
rotation is very significant.

\subsection{Polarization rotation of elliptically polarized
light for large magnetic fields}

\begin{figure}
\centering
\includegraphics[width=1.00\columnwidth]{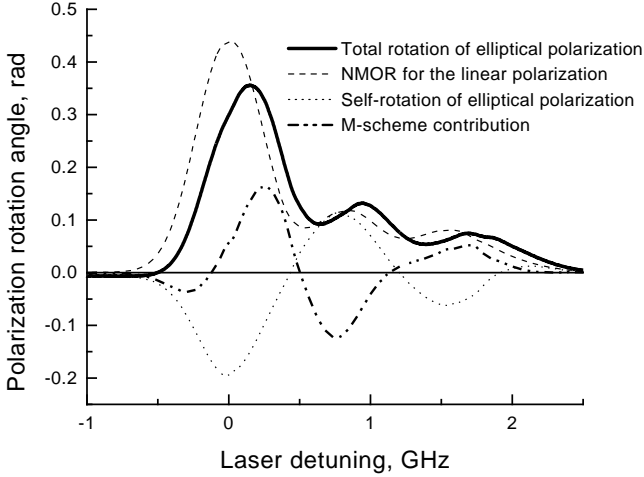}
\caption{\label{rot1.fig}
	The polarization rotation angle as a function of
	laser detuning for ellipticity $\epsilon=25^o$ and
	magnetic field $B=0.35~\mathrm{G}$. The components of
	the rotation due to various processes are also shown.
	Zero detuning corresponds to the $F=2 \rightarrow
	F'=1$ transition. The small peak on the right is due
	to contamination of the cell with ${}^{85}$Rb isotope.
}
\end{figure}

	Now let us consider the case of large magnetic fields.
If the laser frequency is swept across the atomic transition,
the following effects contribute to the polarization rotation:
nonlinear Faraday rotation due to the $\Lambda$-scheme
(experimentally measured for linear polarization), self-rotation
of elliptical polarization due to ac-Stark shifts, and the
magneto-optic rotation of elliptical polarization due to
$M$-scheme induced coherence.  All these components are shown
on Fig.~\ref{rot1.fig}.  It is important to point out that
this ``new'' rotation is comparable with the polarization
rotation for the linear polarization and the self-rotation,
even though this effect is due to higher order nonlinearity.
This proves the effectiveness of the $M$ level scheme for the
enhancement of nonlinear susceptibility in atomic media.

\begin{figure}
\centering
\includegraphics[width=0.70\columnwidth]{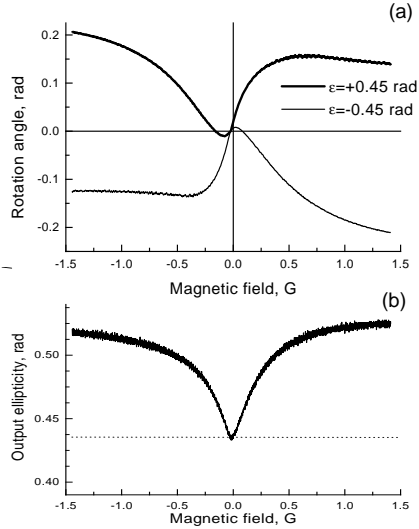}
\caption{\label{rot2.fig}
	(a) The polarization rotation angle as a function of
	magnetic field for opposite values of ellipticity.
	(b) The ellipticity of the transmitted light as a
	function of magnetic field. Initial ellipticity is
	shown as a dashed line.
}
\end{figure}

	The magnetic field dependence of the rotation due
to the ``$M$-scheme'' ground-state coherence reveals a very
peculiar behavior.  When the rotation is independent of the
sign of the ellipticity in the vicinity of zero magnetic field
(as it was demonstrated earlier), then for larger magnetic
fields the rotation becomes asymmetric with respect to both
magnetic field and ellipticity.  To invert the sign of the
rotation, both the ellipticity and the magnetic field should
change their signs (Fig.~\ref{rot2.fig}a).  The ellipticity
of the outgoing light also changes with the magnetic field;
although it is equal to the initial ellipticity for small
magnetic fields (at least for optically thin samples), it
grows symmetrically when the magnetic field becomes larger
(Fig.~\ref{rot2.fig}a).  These changes must to be taken into
account when the polarization rotation angle is measured.

\subsection{NMOR for atoms with higher angular momentum}

	As discussed in Sec.~IV, higher orders of nonlinear
susceptibility may be enhanced in multi-$\Lambda$ systems.
In practice this means that atoms with larger ground state
angular momentum are required.  The most convenient candidate
for the study of the higher orders of Zeeman coherence is
the ${}^{85}\mathrm{Rb}$ isotope, since the same laser may
be used as for our previous study of ${}^{87}\mathrm{Rb}$.
In our experiments we use the $5S_{5/2} F=3 \rightarrow 5P_{3/2}
F'=2$ of ${}^{85}$Rb.  The interaction scheme of elliptically
polarized light with this transition consists of an $M$ scheme
and a triple-$\Lambda$ scheme.

\begin{figure}
\centering
\includegraphics[width=1.00\columnwidth]{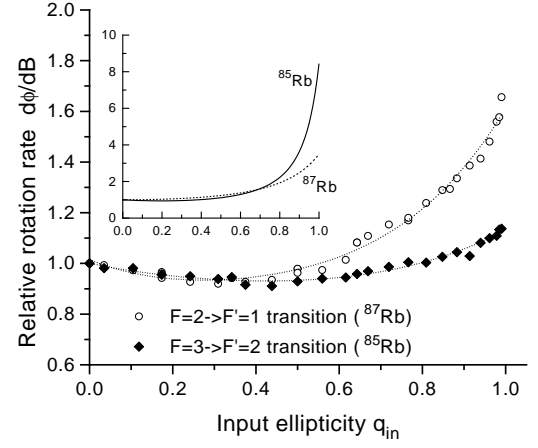}
\caption{ \label{dfidB_iso.fig}
	The normalized slope of nonlinear magneto-optic rotation
	as a function of the ellipticity of the incident
	light for the $F=3 \rightarrow F'=2$ transition of
	${}^{85}$Rb (diamonds), and for $F=2 \rightarrow F'=1$
	transition of ${}^{87}$Rb (circles).  Input laser
	power is P=$2$~mW, the atomic densities are chosen to
	provide $85 \%$ absorption on each transition.  Absolute
	values of the nonlinear Faraday rotation for linear
	polarization were $d\phi/dB(B=0)=2.9~\mathrm{rad/G}$
	and $4.5~\mathrm{rad/G}$ respectively.  Inset: the
	theoretical dependences for naturally broadened
	Rb isotopes, from Eqs.~(\ref{M-rotation}) and
	(\ref{rot-prop-mL}).
}
\end{figure}

	The relative rotation rate for this transition
as a function of the light ellipticity is shown in
Fig.~\ref{dfidB_iso.fig}.  The polarization rotation
enhancement, observed in this case is noticeably smaller
than for ${}^{87}$Rb.  The reason for this may be the smaller
hyperfine splitting of the excited state ($362$~MHz vs $812$~MHz
for ${}^{87}$Rb), which is completely overlapped by the Doppler
broadening ($\Delta_{Doppler} \approx 500$~MHz).  This overlap
results in efficient ``mixing'' of the coherences induced
through different excited states, which may significantly change
the properties of the system.  That is why it would be very
interesting to measure the rotation due to high order coherence,
discussed above, in a cloud of cold atoms. In this case we
expect to see a much stronger effect (Eq.~(\ref{rot-prop-mL})),
since all problems caused by the overlapping transitions due
to the motion of the atoms would be eliminated in cold gas.

\begin{figure}
\centering
\includegraphics[width=1.00\columnwidth]{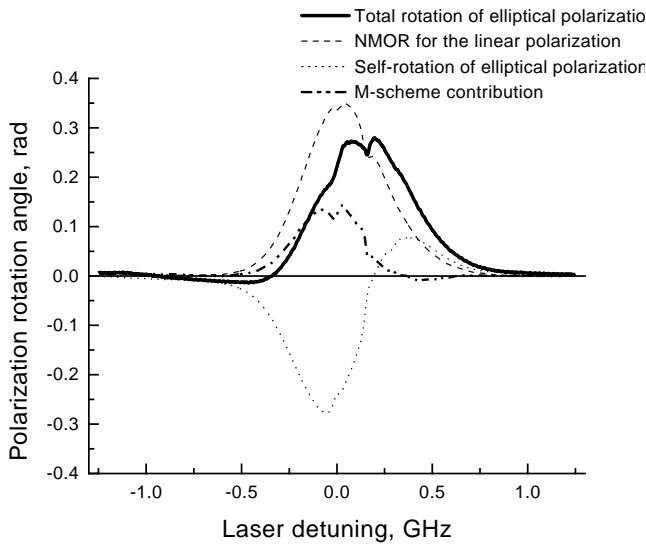}
\caption{ \label{rot1_85.fig}
	The polarization rotation angle in ${}^{85}$Rb
	as a function of laser detuning for ellipticity
	$\epsilon=25^o$ and magnetic field $B=0.35G$. The
	components of the rotation due to various processes
	are also shown.  Zero detuning corresponds to the
	cross-resonance $F=3 \rightarrow F'=2.3$ transition.
	The distortions of the resonances are due to reflected
	light beam.
}
\end{figure}

	The spectral dependence of the rotation of the
elliptical polarization on laser frequency for the case of large
magnetic field is shown in Fig.~\ref{rot1_85.fig}.  Similarly to
the ${}^{87}$Rb, the high-order Zeeman coherence significantly
modifies the rotation spectra, and the contribution of the
nonlinear rotation is comparable with the rotation of the
linear polarization and self-rotation.

	One can see additional sub-Doppler structure on top
of the rotation resonances.  These peaks appears due to the
retro-reflection of the laser beam inside the atomic cell.
This additional beam interacts with atoms and causes the
redistribution of the atomic population similar to Doppler-free
saturation spectroscopy.

\section{Conclusion}

	We have studied the nonlinear magneto-optic rotation
of elliptically polarized light interacting with various
transitions of rubidium atoms.  We have shown that this rotation
can be described by means of $\Lambda$, $M$, and higher chain
$\Lambda$ schemes.  For the simple three-level $\Lambda$
scheme, the rotation does not depend on the light ellipticity.
For more complicated systems, the multi-photon processes
are responsible for the creation of high-order ground-state
coherence resulting in a new type of ellipticity-dependent
nonlinear magneto-optical rotation.  We have derived simple
analytical expressions for this rotation for the $M$ interaction
scheme (Eq.~(\ref{M-rotation})) and we showed that this effect
results from the coherently induced hexadecapole moment.

	Since the modification of NMOR is associated with an
enhancement of nonlinear atomic susceptibility, we have analyzed
the effectiveness of this process by comparing the nonlinear
susceptibility for $M$ and $N$ interaction schemes.  We have
demonstrated that although the enhancements of nonlinearity in
these schemes are caused by different mechanisms, they exhibit
the same absorptive and refractive nonlinearity magnitudes.
We have also shown that the generalized $M$ scheme may be
used to create resonantly enhanced nonlinear susceptibility
of any given order, similarly to the generalized $N$
scheme~\cite{zubairy02pra}.  We have discussed the possible
implementation of the generalized $M$ scheme for quantum
computer algorithms.

	To verify our theoretical calculations, we have
studied the polarization rotation of elliptically polarized
laser light propagating through Rb vapor.  The $M$ interaction
scheme is realized on the $F=2 \rightarrow F'=1$ transition of
${}^{87}$Rb, and the triple-$\Lambda$ scheme is observed on the
$F=3 \rightarrow F'=2$ transition of ${}^{87}$Rb. Although the
experimental points cannot be fit perfectly by the theoretical
expressions (Eqs.~(\ref{lambda-rot}) and (\ref{M-rotation})),
the basic properties of the new rotation are confirmed.

\begin{acknowledgments}

	The authors  gratefully acknowledge useful discussions
with  D.\ Budker, A.\ D.\ Greentree, and M.\ O.\ Scully, and
the support from Air Force Research Laboratory, DARPA-QuIST,
TAMU Telecommunication and Informatics Task Force (TITF)
initiative, and the Office of Naval Research.

\end{acknowledgments}



\printfigures


\begin{thebibliography}{99}
\bibitem{note1}
This and following discussions are not applicable to the case of
pure circularly polarized light, since the definition of the
polarization rotation is not valid then.
\bibitem{note}
We consistently use two definitions for the light ellipticity: the
relative difference of the circular component intensities ($q$),
and the phase delay between two linear components ($\epsilon$).
They are related to each other the following way: $q=\sin
2\epsilon$.
\end{thebibliography}
\end{document}